\documentclass[twocolumn,showpacs,preprintnumbers,amsmath,amssymb,superscriptaddress]{revtex4-1}

\usepackage{amssymb,bm}
\usepackage{booktabs}
\usepackage{graphicx,color}
\newcommand{\mgdel}[1]{}

\newcommand{\lmdel}[1]{} 
\newcommand{\lton}{\mathrel{\lower.9ex 
  \hbox{$\stackrel{\displaystyle <}{\sim}$}}}

\newcommand{\ber}{\begin{eqnarray}}
\newcommand{\eer}{\end{eqnarray}}

\newcommand{\beq}{\begin{equation}}
\newcommand{\eeq}[1]{\label{#1} \end{equation}}

\newcommand{\ee}{\end{equation}} \newcommand{\ben}{\begin{enumerate}}
\newcommand{\een}{\end{enumerate}} \newcommand{\bit}{\begin{itemize}}
\newcommand{\eit}{\end{itemize}} \newcommand{\bc}{\begin{center}}
\newcommand{\ec}{\end{center}} \newcommand{\bea}{\begin{eqnarray}}
\newcommand{\eea}{\end{eqnarray}}
\newcommand{\beqar}{\begin{eqnarray}}
\newcommand{\eeqar}[1]{\label{#1} \end{eqnarray}}

\def\ben{\begin{eqnarray}}
\def\enn{\end{eqnarray}}

\def\bk{{\bf{k}}}

\def\bq{{\bf{q}}}

\def\be{\begin{equation}}
\def\ee{\end{equation}}
\def\bea{\begin{eqnarray}}
\def\eea{\end{eqnarray}}

\newcommand{\hatq}{\hat{q}}
\newcommand{\Pq}{P_{q(\bar{q}})}
\newcommand{\Pg}{P_g}
\newcommand{\bfqperp}{{\bf q}_\perp}
\newcommand{\ave}[1]{\langle {#1} \rangle}

\begin{document}

\title{Extracting jet transport coefficient from jet quenching at RHIC and LHC}


\author{Karen M. Burke}
\affiliation{Department of Physics and Astronomy, Wayne State University, Detroit, Michigan 48201, USA}

\author{Alessandro Buzzatti}
\affiliation{Nuclear Science Division, MS 70R0319, Lawrence Berkeley National Laboratory, Berkeley, California 94720, USA}
\affiliation{Department of Physics, Columbia University, New York 10027, USA}

\author{Ningbo Chang}

\affiliation{School of Physics, Shandong University, Jinan, Shandong 250100, China}
\affiliation{Institute of Particle Physics and Key Laboratory of Quarks and Lepton Physics (MOE), Central China Normal University, Wuhan 430079, China}

\author{Charles Gale}
\affiliation{Department of Physics, McGill University, 3600 University Street, Montreal, Quebec, H3A 2T8, Canada}

\author{Miklos Gyulassy}
\affiliation{Department of Physics, Columbia University, New York 10027, USA}

\author{Ulrich Heinz}
\affiliation{Department of Physics, The Ohio State University, Ohio 43210, USA}

\author{Sangyong Jeon}
\affiliation{Department of Physics, McGill University, 3600 University Street, Montreal, Quebec, H3A 2T8, Canada}

\author{Abhijit Majumder}
\affiliation{Department of Physics and Astronomy, Wayne State University, Detroit, Michigan 48201, USA}

\author{Berndt M\" uller}
\affiliation{Department of Physics, Brookhaven National Laboratory, Upton, New York 11973, USA}

\author{Guang-You Qin}
\affiliation{Institute of Particle Physics and Key Laboratory of Quarks and Lepton Physics (MOE), Central China Normal University, Wuhan 430079, China}
\affiliation{Department of Physics and Astronomy, Wayne State University, Detroit, Michigan 48201, USA}

\author{Bj\"orn Schenke}
\affiliation{Department of Physics, Brookhaven National Laboratory, Upton, New York 11973, USA}

\author{Chun Shen}
\affiliation{Department of Physics, The Ohio State University, Ohio 43210, USA}

\author{Xin-Nian Wang}
\affiliation{Institute of Particle Physics and Key Laboratory of Quarks and Lepton Physics (MOE), Central China Normal University, Wuhan 430079, China}
\affiliation{Nuclear Science Division, MS 70R0319, Lawrence Berkeley National Laboratory, Berkeley, California 94720, USA}

\author{Jiechen Xu}
\affiliation{Department of Physics, Columbia University, New York 10027, USA}

\author{Clint Young}
\affiliation{School of Physics and Astronomy, University of Minnesota, Minneapolis, Minnesota  55455, USA}

\author{Hanzhong Zhang}
\affiliation{Institute of Particle Physics and Key Laboratory of Quarks and Lepton Physics (MOE), Central China Normal University, Wuhan 430079, China}



\begin{abstract}	
	
	\centerline{(The JET Collaboration)}
	
\vspace{12pt}

	Within five different approaches to parton propagation and energy loss in dense matter, a phenomenological study of experimental data on suppression of large $p_T$ single inclusive hadrons in heavy-ion collisions at both RHIC and LHC was carried out.  The evolution of bulk medium used in the study for parton propagation was given by 2+1D or 3+1D hydrodynamic models which are also constrained by experimental data on bulk hadron spectra. Values for the jet transport parameter $\hat q$ at the center of the most central heavy-ion collisions are extracted or calculated within each model, with parameters for the medium properties that are constrained by experimental data on the hadron suppression factor $R_{AA}$.   For a quark with initial energy of 10 GeV we find that $\hat q\approx 1.2 \pm 0.3$ GeV$^2$/fm at an initial time $\tau_0=0.6$ fm/$c$ in Au+Au collisions at $\sqrt{s}=200$ GeV/n and $\hat q\approx 1.9 \pm 0.7 $ GeV$^2$/fm in Pb+Pb collisions at $\sqrt{s}=2.76 $ TeV/n. Compared to earlier studies, these represent significant convergence on values of the extracted jet transport parameter, reflecting recent advances in theory and the availability of new experiment data from the LHC.
	
\end{abstract}

\maketitle

\section{Introduction}

In the search and study of the quark-gluon plasma (QGP)  in high-energy heavy-ion collisions, jet quenching processes
play an essential role as hard probes of the properties of dense matter. Because of the hard scales involved, jets are produced
in the very early stage of the collisions and their initial production rate can be calculated within perturbative QCD. During their
subsequent propagation through the dense medium, interaction between jets and medium will lead to jet energy loss and suppression
of final jets and large transverse momentum hadron spectra. Original theoretical 
studies based on this principle \cite{Bjorken:1982tu}--\cite{Gyulassy:2000gk} and collaborative work by the Hard Probes 
Collaboration on the survey of hard processes in the absence of a hot or dense QCD medium \cite{hpc1,hpc2} formed 
the basis for the initial success of the RHIC experimental program on hard probes and the phenomenological 
studies that ensued.

Since the start of the Relativistic Heavy-ion Collider (RHIC) experimental program, we have seen not only the suppression of 
single inclusive hadron spectra at large transverse momentum \cite{star-suppression,phenix-suppression} but also of back-to-back
high $p_{T}$ dihadron \cite{stardihadron} and $\gamma$-hadron correlations \cite{Wang:1996yh,Adare:2009vd,Abelev:2009gu}.  
The same jet quenching patterns are also observed in the latest heavy-ion collisions at the Large Hadron Collider (LHC)
\cite{CMS:2012aa, Abelev:2012hxa,Aamodt:2011vg}. In addition, one has also observed the predicted suppression of reconstructed jets \cite{Vitev:2009rd,Aad:2012vca,Aad:2013sla},  as well as increased dijet \cite{Aad:2010bu,Chatrchyan:2012nia}  and $\gamma$-jet asymmetry \cite{Chatrchyan:2012gt,ATLAS:2012cna}. These observed jet quenching phenomena in heavy-ion collisions at RHIC 
have been studied within a variety of models \cite{Wang:2004dn,Vitev:2002pf,Wang:2003mm,Eskola:2004cr,Renk:2006nd,Zhang:2007ja,Qin:2007rn,Renk:2006qg, Zhang:2009rn,Qin:2009bk,Gyulassy:2003mc,Kovner:2003zj,Majumder:2010qh} that incorporate parton energy loss as jets propagate through dense matter.  Though many models can describe the observed jet quenching at RHIC quite well by adjusting parameters, there exist differences in the implementation of parton energy loss in jet quenching models \cite{Armesto:2011ht}. New data from LHC experiments have lent support to some of these models  \cite{Chen:2011vt,Majumder:2011uk,Zapp:2012ak} while  challenging 
others \cite{CMS:2012aa,Abelev:2012hxa,Renk:2013rla}.  Even within those models that can describe experimental data, the combined data from experiments at RHIC and LHC provide unprecedented constraints on the medium parameters as probed by jet quenching. 

One of the programmatic goals of heavy-ion collisions is to extract important medium properties from phenomenological studies of combined experimental data on a wide variety of jet quenching measurements. This is also one of the goals of the JET Collaboration.  As a first step toward this goal we carry out in this paper a survey study of medium properties within some of the existing approaches to medium-induced parton energy loss, using constraints provided by experimental data on suppression of large transverse momentum single inclusive hadron spectra at RHIC and LHC. We will assess five different approaches to parton energy loss: GLV-CUJET, HT-M, HT-BW, MARTINI and McGill-AMY.  GLV \cite{Gyulassy:2000er4} and its recent CUJET implementation \cite{Buzzatti:2011vt} use a potential model for multiple scattering in the medium in which the controlling parameters for energy loss are the strong coupling constant, the Debye screening mass and the density of scattering centers. Within the high-twist (HT) approaches (HT-BW and HT-M)\cite{Chen:2011vt,Majumder:2011uk},  the jet transport coefficient or averaged transverse momentum squared per unit length is the only medium property that affects the parton energy loss. The MARTINI \cite{Schenke:2009gb} and McGill-AMY \cite{Qin:2007rn} model are based on hard-thermal-loop (HTL) resummed thermal field theory in which the only adjustable parameter is the strong coupling constant.  To have a common ground for this survey study, we focus on the jet transport coefficient $\hat{q}$ for a jet initiated by a light quark as given by each of the parton energy loss models. While in the HT approaches this is a direct fit parameter, in the other approaches it can be computed from the respective fitted model parameters.

Since the energy loss or medium modification of the final hadron spectra depends on the space-time profile of  parton density in the medium, any systematic and qualitative extraction of the properties of the medium through phenomenological study of jet quenching has to take into account the dynamical evolution of the bulk matter \cite{Bass:2008rv,Armesto:2009zi,Chen:2010te}. For our current study,  2+1D \cite{Song:2007fn,Song:2007ux,Qiu:2011hf,Qiu:2012uy}  or 3+1D \cite{Hirano2001,HT2002,Nonaka:2006yn,Schenke:2010nt,Schenke:2010rr}  ideal or viscous hydrodynamic simulations provide the most realistic dynamic evolution of the bulk medium available that are constrained by experimental data on bulk hadron production, including charged hadron spectra and their azimuthal anisotropies.  Here we use event averaged initial conditions for the bulk matter evolution. Uncertainties in jet quenching calculations as a result of variations in hydrodynamic bulk evolution due to event-by-event initial conditions and associated changes in the value of the shear viscosity are expected to be small once they are constrained by the experimental data on bulk hadron productions in heavy-ion collisions. 

Similar efforts to extract values of the jet quenching parameter have been made before \cite{Bass:2008rv,Armesto:2009zi} but with diverging values from different models varying as much as a factor of 8.  Our present work will take advantage of the significant progress in our theoretical understanding and modeling of jet quenching and of the evolving medium created in heavy-ion collisions at RHIC and LHC. For the first time in such a comprehensive study, we evaluate the range of jet transport parameters allowed by the combined experimental data at RHIC and LHC.  
As we shall see, the availability of new data on heavy-ion collisions at LHC,  where higher initial temperature is reached and the range of $p_T$ is much larger than at RHIC,  allows us to investigate the temperature and jet energy dependence of the jet transport coefficient. 

In the remainder of this paper, we will review briefly in Secs. II-VI the five different approaches to parton energy loss employed in this work.  We investigate constraints on the jet transport parameter in each model by comparing the calculated suppressions factors for single hadron spectra with the experimental data at RHIC and LHC.
We compile these constraints in Sec. VII to provide an up-to-date estimate of the jet transport parameter and its temperature dependence within the range that has been reached in the most central Au+Au collisions at RHIC and Pb+Pb collisions at LHC. A summary and discussions are given in Sec. VII.

\section{GLV-CUJET model}

The GLV model \cite{Vitev:2002pf,Gyulassy:2000er4} correctly predicted in 2002 the general form of the 
$\sqrt{s}$ evolution of the high $p_T$ pion  nuclear modification factor,
\begin{equation}
R_{AA}(p_T,\eta=0;\sqrt{s}, b)= \frac{dN_{AA\rightarrow \pi}/d^2p_T}{T_{AA}(b)d\sigma_{pp\rightarrow \pi}/d^2p_T},
\end{equation}
from SPS,RHIC to LHC energies.
GLV was generalized to include thermal mass and heavy quark
effect in DGLV\cite{Djordjevic:2004}.
However in 2005 PHENIX discovered  DGLV significantly under-predicted 
quenching of charm and bottom quark jets. This led to the 
WHDG\cite{Wicks:2005gt01} generalization of DGLV\cite{Djordjevic:2004} theory
to check whether quenching effects due to  elastic energy loss 
and more realistic jet path length fluctuations could account for the 
non-photonic electron spectrum from heavy quark meson decay
data from PHENIX at RHIC. We found that those effects did not solve the ``heavy quark jet puzzle''.  This led to the dynamical  generalization
of DGLV\cite{Djordjevic09} replacing the GW static color electric scattering center into the Hard Thermal Loop (HTL) weakly coupled Quark Gluon Plasma 
ansatz. The jet medium interactions with a HTL QGP  medium 
include  dynamic color magnetic  as well as   static color electric 
interactions. 

The CUJET1.0 Monte Carlo code was developed at Columbia University as part of the Topical JET Collaboration project. 
With this code we were able to predict the full jet quenching pattern for both light ($\pi$) and heavy flavor (D and B) hadrons at both 
RHIC and LHC including dynamical DGLV, elastic energy loss,  as well as full space+time evolution background of the HTL QGP bulk medium. 
The CUJET1.0 code featured: 
\begin{enumerate}
\item a dynamical jet interaction potentials 
that can interpolate between pure HTL dynamically screened magnetic and static electric screening limits;
\item the ability to calculate high order opacity corrections up to 9th order in opacity; 
\item integration over jet path in diffuse nuclear geometries including Bjorken longitudinally expanding HTL QGP;
\item inclusion of local multi-scale running coupling effects for radiative energy loss and flexibility to explore nonperturbative deformations of HTL screening scales;
\item inclusion of running coupling elastic energy loss with fluctuations;
\item convolution over initial jet spectra from pQCD parton models; and
\item convolution over jet fragmentation functions and semileptonic final decay into non-photonic electrons.
\end{enumerate}

CUJET1.0 succeeded in explaining for the first time \cite{Buzzatti:2011vt}
 the  anomalous high quenching of non-photonic electrons within a pure HTL
QCD paradigm and thus provided a natural solution 
to the old heavy quark jet puzzle at RHIC as due to
 enhanced dynamical magnetic scattering effects. 
It further predicted a novel inversion of the $\pi < D  < B$  flavor ordering of $R_{AA}$ at
high $p_T$ that can be tested in the future at RHIC and LHC . 

One of the surprising \cite{Horowitz:2011gd} LHC discoveries was the
similarity between $R_{AA}$ at RHIC and LHC despite the doubling of the initial QGP
density from RHIC to LHC.  CUJET1.0 was able to  explain
this by taking into account the effects due to multi-scale running of the QCD
coupling $\alpha(Q^2)$ in the DGLV opacity series. At first order in
opacity the running coupling rcDGLV induced gluon radiative
distribution is given by \cite{Buzzatti:2012dy}
\begin{widetext}
\begin{eqnarray}
x\frac{dN_{Q\rightarrow Q+g}}{dx}({\bf r},\phi) &= &
 \int d\tau \rho({\bf r}+\hat{{\bf n}}(\phi)\tau, \tau)  \int  \frac{d^2{\bq}_T}{\pi} \frac{d^2\sigma_{\rm eff}}{d^2{\bq}_T} \int \frac{d^2{\bk}_T}{\pi} 
\alpha_{\rm s}(k_T^2/(x(1-x))
  \nonumber \\
&\times& \frac{12(\bk_T{+}\bq_T)}{(\bk_T{+}\bq_T)^2{+}\chi(\tau)}
	\cdot \left( \frac{(\bk_T{+}\bq_T)}{(\bk_T{+}\bq_T)^2{+}\chi(\tau)} - \frac{\bk_T}{\bk^2_T{+}\chi(\tau)} \right) 
\left(  1- \cos\left[\frac{(\bk_T{+}\bq_T)^2+\chi(\tau)}{2x_+E}\; \tau\right] \right) ,
\end{eqnarray}
\end{widetext}
where the effective running differential quark-gluon cross section is
\begin{eqnarray}
\frac{d^2\sigma_{\rm eff}}{d^2\bq_T}=\frac{\alpha^2_{\rm s} (\bq^2_T)}{(\bq^2_T{+}f^2_E \mu^2(\tau))(\bq^2_T{+}f^2_M \mu^2(\tau))}\;,
\end{eqnarray}
that runs with both $q_T$ and the local temperature through
$\mu^2(\tau)=4\pi \alpha_{\rm s}(4T^2) T^2$, the local HTL color
electric Debye screening mass squared in a pure gluonic plasma with local
temperature $T(\tau)\propto \rho^{1/3} ({\bf r},\tau)$ along 
the jet path ${\bf r}(\tau)$ through the plasma. 
 
Here the infrared scale $\chi(\tau)=M^2 x_+^2 + f_E^2 \mu^2(T(\tau)) (1-x_+)/\surd{2}$ 
controls the ``dead cone'' and  LPM destructive interference 
effects due to both the finite quark current mass $M$,
 and a asymptotic thermal gluon mass assumed of the form 
$m_g=f_E\, \mu(T)/\sqrt{2}$ mass.

The HTL deformation parameters $(f_E,f_M)$ are used to vary the electric and
magnetic screening scales relative to HTL.  In general HTL deformations
could also change $m_g(T)$. The default HTL plasma  is
$(1,0)$ but we also consider a deformed $(2,2)$ plasma model motivated by
lattice QCD screening data.  The vacuum running $\alpha_{\rm s}(Q^2) =
\min[\alpha_{max}, 2\pi/9\log(Q^2/\Lambda^2)]$ is used which is characterized 
by a nonperturbative  maximum value $\alpha_{max}$. The
parameters $(\alpha_{max}, f_E, f_M)$ are therefore the main model
control parameters in this study.

The computational task performed via Monte Carlo integration is to
evaluate $dN_g/dx$ for each $({\bf r},\hat{n})$ initial jet production
coordinates, convolute the inclusive gluon spectrum via a Poisson ansatz
to estimate effects of multi-gluon fluctuation, 
evaluate the normalized radiation probability,
$P_{rad}(\Delta E_{rad}, E_0; {\bf r},\hat{n}))$ via fast Fourier
transform including delta function $\Delta E/E_0 =0,1$ end point sigularities. 
Multiple running coupling elastic energy
loss probability, $P_{el}(\Delta E_{el}, E_0; {\bf r},\hat{n}))$ is computed,
and then convoluted $P_{rad}\otimes P_{el}$ with probability for radiative energy loss. The final
total energy loss probability is then folded over the initial parton jet
spectrum $dN_{pp}/d^2p_Td\eta$.  Finally CUJET averages over initial jet
configurations via $\int d^2{\bf r} d^2\hat{n} T_A({\bf r}+ {\bf b}/2)
T_A({\bf r},{\bf b}/2)$ and fragments jets into different flavor
hadrons or leptons to compare with data.

In CUJET2.0, CUJET1.0 model is coupled to state of the art 2+1 D viscous hydro fields with shear viscosity to entropy density ratio $\eta/s=0.08$ 
\cite{Shen:2010uy,Shen:2011eg} as tabulated by 
 the hydro group within the JET Collaboration.  
The hydro temperature fields used in CUJET2.0 \cite{JXMG13}
 are thus constrained by thermal and flow fields that
fit experimental data on bulk low $p_T<2$ GeV/$c$ radial and elliptic flow
observables.  The effects of azimuthally
asymmetric radial flowing QGP are then be computed via 
the  CUJET2.0=rcDGLV+VISH  C++ code.

\begin{figure}
\centering
\includegraphics[width=3.5in ]{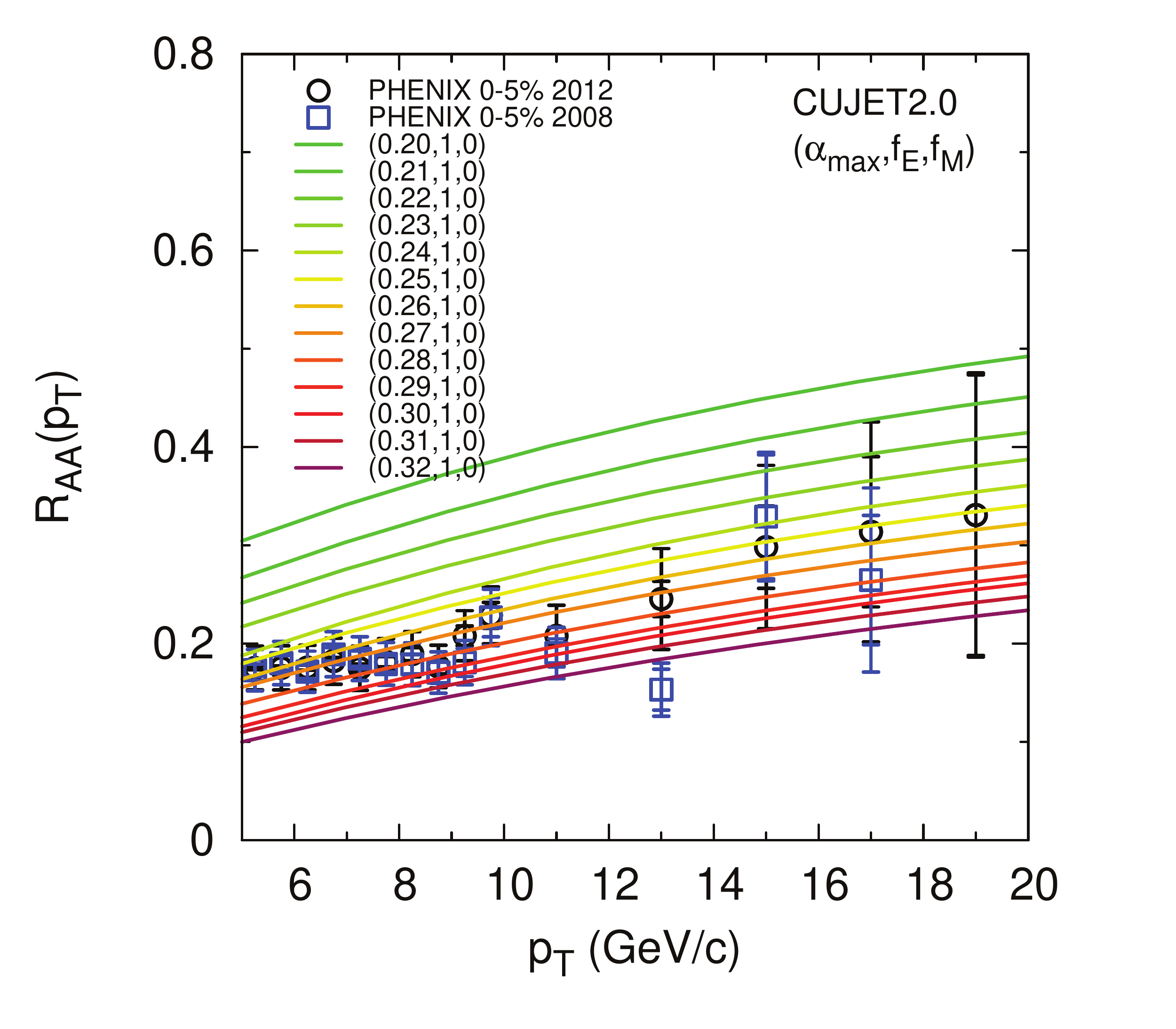}

\vspace{-0.6 cm}
\includegraphics[width=3.5in ]{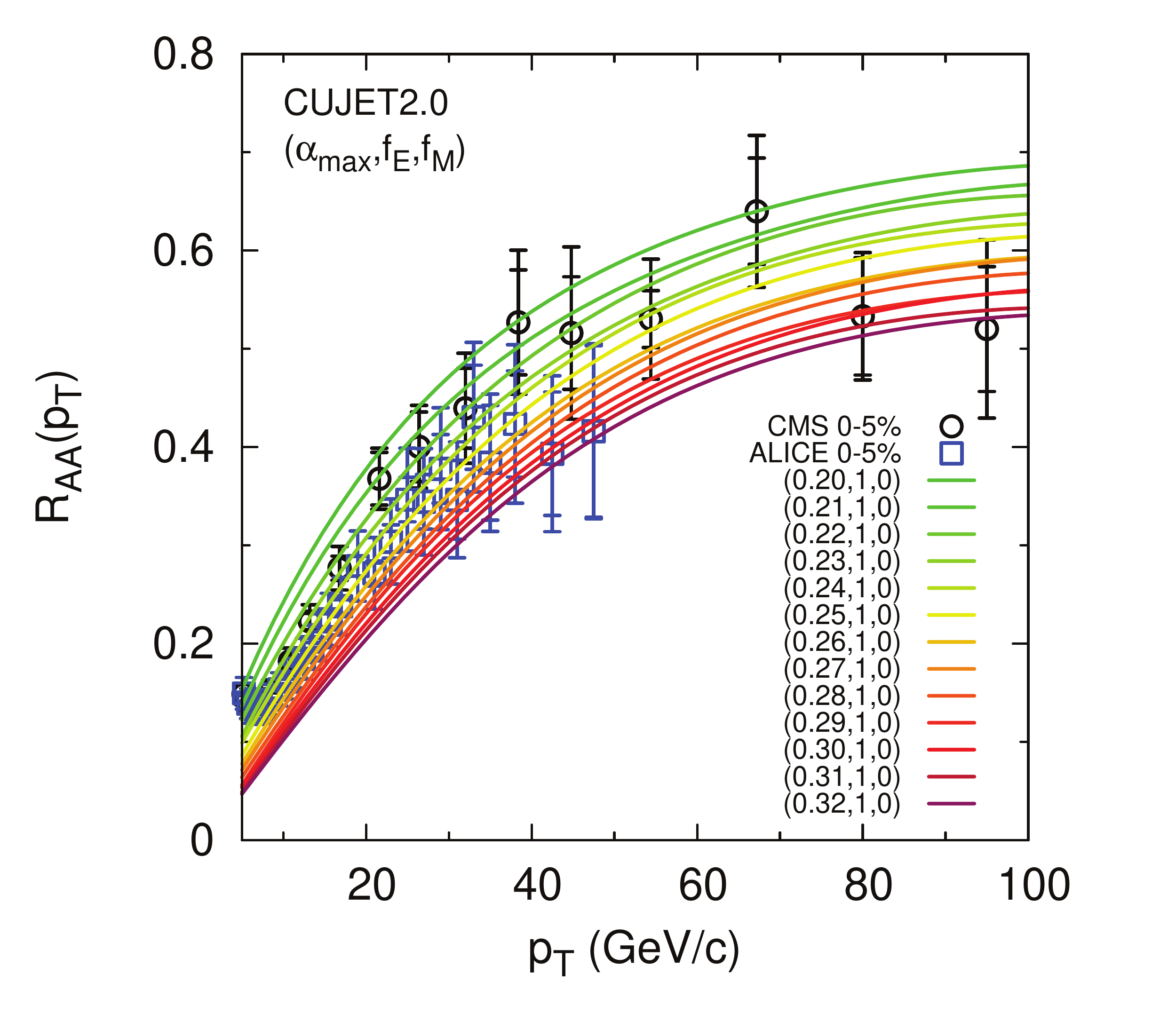}
\caption{(Color online) CUJET results for the nuclear modification factor at mid-rapidity for neutral pion spectra in  $0-5\%$ central Au+Au collisions at $\sqrt{s}=200$ GeV/n (upper panel) and for charged hadrons in Pb+Pb collisions at $\sqrt{s}=2.76$ TeV/n (lower panel) 
with a range of values of frozen strong coupling constant $\alpha_{\rm max}$, as compared to PHENIX data \cite{Adare:2008qa,Adare:2012wg} at RHIC and ALICE \cite{Abelev:2012hxa} and CMS data \cite{CMS:2012aa} at LHC.} 
\label{fig:glv-raa}
\end{figure}

\begin{figure}
\includegraphics[width=3.5in ]{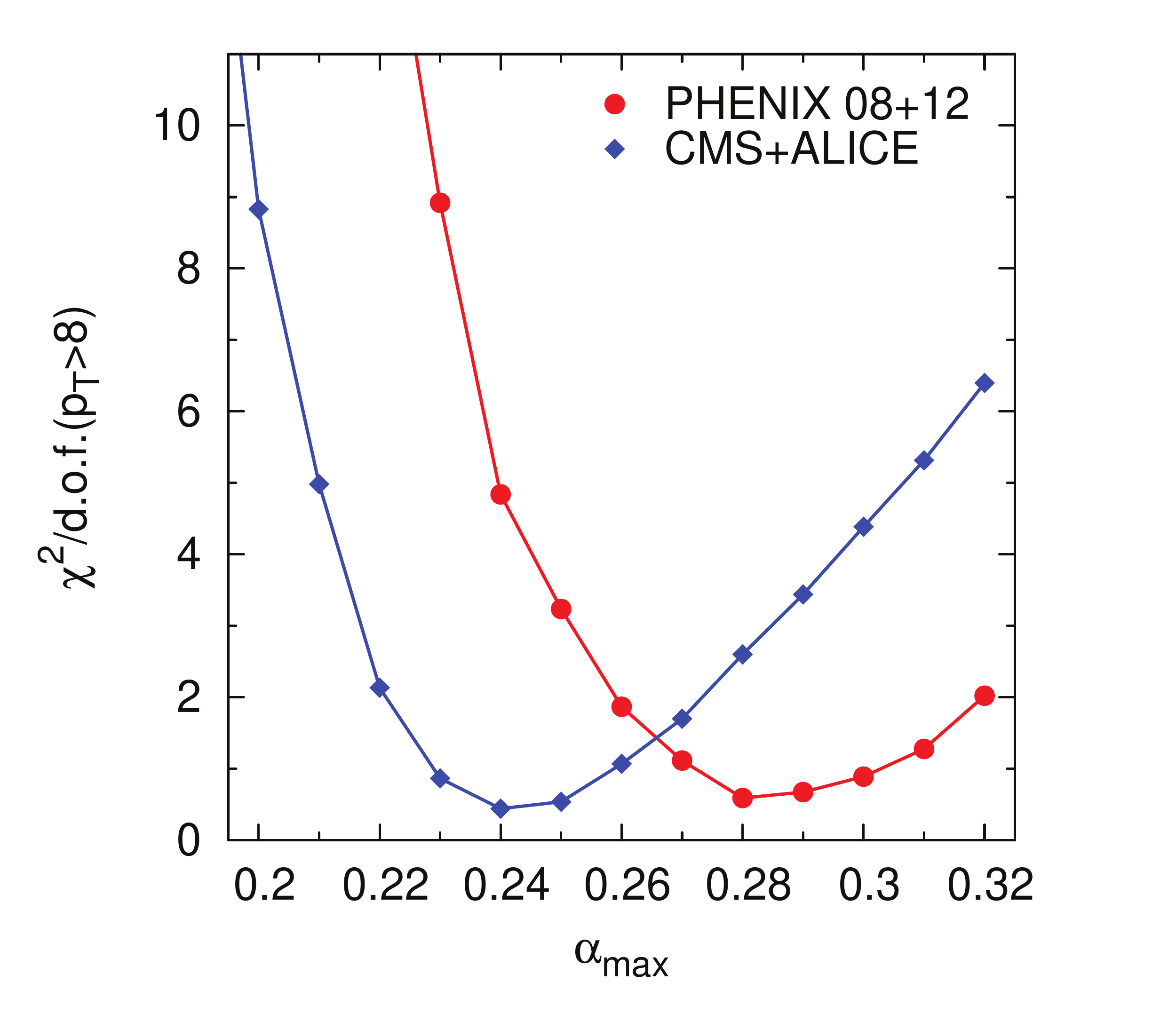}
\caption{ (Color online) The $\chi^2$/d.o.f as a function of the model parameter $\alpha_{\rm max}$ from fitting to the PHENIX data \cite{Adare:2008qa,Adare:2012wg}  (combined 2008 and 2012 data set) at RHIC and combined ALICE \cite{Abelev:2012hxa} and CMS \cite{CMS:2012aa}  data at LHC by the CUJET model
calculation of the nuclear suppression factor $R_{AA}(p_T)$ as shown in Fig.~\ref{fig:glv-raa}.
\label{fig:glv-chi2} } 
\end{figure}

Shown in Fig.~\ref{fig:glv-raa} are the calculated single hadron suppression factor $R_{AA}(p_T)$ for central Au+Au collisions at RHIC
and Pb+Pb collisions at LHC with a range of parameters $(\alpha_{max}, f_E, f_M)$  = $(\alpha_{max}, 1, 0)$ as compared to experimental data. The $\chi^2$/d.o.f from fits to the experimental data as a function of $\alpha_{max}$ are shown in Fig.~\ref{fig:glv-chi2}. The
experimental data at RHIC and LHC seems to prefer different values of  $\alpha_{max}$. One can consider the 
range $\alpha_{max}=0.22-0.31$ as the systematic uncertainty of the model parameter via fits to experimental data at RHIC and LHC.
In the future the sensitivity to varying the running coupling scales will also be investigated. 

The physics implications of these solutions can be visualized by computing the effective
jet transport coefficient 
\begin{equation}
\hat{q}(E,T; \alpha_{max},f_E,f_M)=\rho_g(T)\int_0^{\sqrt{6ET}}
dq^2_T q^2_T \frac{d\sigma_{qg}}{dq^2_T}
\label{eq:qhat-tmb}
\end{equation}
in an idealized static and homogeneous thermal equilibrium
medium. This jet transport coefficient $\hat{q}$ depends on jet energy $E$ and 
temperature $T$ variations that  also influence the fitted values of  
$\alpha_{max}$ as well 
as well as electric and magnetic screening mass deformations parameters
$(f_E,f_M)$. We have found e.g., that the default HTL model $(1,0)$
has lower $\chi^2$ than the $(2,0)$ deformed HTL model. We will search 
in the future  for the global minimum $\chi^2(\alpha_{max},f_E,f_M)$ that best 
fits the combined RHIC and LHC data on the centrality dependence of $R_{AA}(p_T,b,\sqrt{s})$ and especially  the jet elliptic moments $v_2(p_T)$ which 
remain especially challenging at this time.

\section{Higher-Twist-Berkeley-Wuhan (HT-BW) model}

Within a high-twist approach (HT) \cite{Guo:2000nz,Wang:2001ifa},
medium-modified quark fragmentation functions are given by
\begin{eqnarray}
\tilde{D}_{q}^{h}(z_h,Q^2) &=&
D_{q}^{h}(z_h,Q^2)+\frac{\alpha_s(Q^2)}{2\pi}
\int_0^{Q^2}\frac{d\ell_T^2}{\ell_T^2} \nonumber\\
&&\hspace{-0.7in}\times \int_{z_h}^{1}\frac{dz}{z} \left[ \Delta\gamma_{q\rightarrow qg}(z,\ell_T^2)D_{q}^h(\frac{z_h}{z})\right.
\nonumber\\
&&\hspace{25pt}+\left. \Delta\gamma_{q\rightarrow
gq}(z,\ell_T^2)D_{g}^h(\frac{z_h}{z}) \right] ,
\label{eq:mo-fragment}
\end{eqnarray}
which take a form very similar to the vacuum bremsstrahlung corrections that lead to the
evolution equations in pQCD for fragmentation
functions, except that the medium modified splitting functions,
$\Delta\gamma_{q\rightarrow qg}(z,\ell_T^2)$ and
$\Delta\gamma_{q\rightarrow gq}(z,\ell_T^2)=\Delta\gamma_{q \rightarrow qg}(1-z,\ell_T^2)$
depend on the properties of the medium via the jet transport
parameter $\hat q$ in Eq.~(\ref{eq:qhat-tmb}), the average squared transverse momentum broadening per unit
length. In the HT approach the jet transport parameter for a quark is related to the gluon distribution density of the
medium \cite{Baier:1996sk,CasalderreySolana:2007sw},
function
\begin{equation}
\hat{q} =\frac{4\pi C_F \alpha_{\rm s}}{N_c^2-1} \int dy^-
\left\langle F^{ai+}(0)F_{i}^{a+}(y^-) \right\rangle e^{i \xi p^+y^-},
\label{eq:qhat-ff}
\end{equation}
where, $\langle {\cal O} \rangle =(2\pi)^{-3} \int d^3p/2p^+ f(p) \langle p| {\cal O} |p\rangle$
denotes the ensemble average of an operator ${\cal O}$ in the medium
composed of states $|p\rangle$ with occupation probability $f(p)$,
$\xi=\langle k_T^2\rangle/2E\langle p^+\rangle$, $\langle k_T^2\rangle$ 
is the average transverse momentum carried by the gluons in $|p\rangle$, and 
$\rho=\int d^3p f(p)/(2\pi)^3$ denotes the density of scattering centers in the matter.

The corresponding quark energy loss can be expressed as \cite{Chen:2010te,benwei-nuleon},
\begin{eqnarray}
\frac{\Delta E}{E} &=& \frac{2N_{c}\alpha_s}{\pi} \int dy^-dz
{d\ell_T^2}
\frac{1+(1-z)^2}{\ell_T^4} 
\nonumber \\ &\times&  \left(1-\frac{1-z}{2}\right)
\hat q(E,y)
\sin^2\left[\frac{y^-\ell_T^2}{4Ez(1-z)}\right], \label{eq:de-twist}
\end{eqnarray}
in terms of the jet transport parameter for a quark jet.  Note that an extra factor of $1-(1-z)/2$ is included here as compared to that
used in Refs. \cite{CasalderreySolana:2007sw,Deng:2009qb} due to corrections beyond the helicity
amplitude approximation \cite{benwei-nuleon}. 

According to the definition of jet transport parameter, we can assume it to be proportional
to local parton density in a QGP and hadron density in a hadronic gas. Therefore, in
a dynamical evolving medium, one can express it in general as \cite{Chen:2010te,Chen:2011vt,CasalderreySolana:2007sw}
\begin{equation}
\label{q-hat-qgph}
\hat{q} (\tau,r)= \left[\hat{q}_0\frac{\rho_{QGP}(\tau,r)}{\rho_{QGP}(\tau_{0},0)}
  (1-f) + \hat q_{h}(\tau,r) f \right]\cdot \frac{p\cdot u}{p_0}\,,
\end{equation}
where $\rho_{QGP}$ is the parton (quarks and gluon) density in an ideal gas at a given temperature,
$f(\tau,r)$ is the fraction of the hadronic phase at any given
space and time, $\hat q_{0}$ denotes the jet transport
parameter for a quark at the center of the bulk medium in the QGP phase at the
initial time $\tau_{0}$, $p^\mu$ is the four momentum of the jet and $u^\mu$ is the
four flow velocity in the collision frame. The hadronic phase of the medium
is assumed to be a hadron resonance gas, in which the jet transport parameter is approximated as,
\begin{equation}
\hat q_{h}=\frac{\hat q_{N}}{\rho_{N}}\left[ \frac{2}{3}\sum_{M}\rho_{M}(T)+\sum_{B}\rho_{B}(T)\right],
\label{eq:qhath}
\end{equation}
where $\rho_{M}$ and $\rho_{B}$ are the meson and baryon density in the hadronic resonance gas at
a given temperature, respectively, $\rho_{N}=n_{0}\approx 0.17$ fm$^{-3}$ is the nucleon density in the
center of a large nucleus and the
factor $2/3$ accounts for the ratio of constituent quark numbers in mesons and baryons.
The jet transport parameter for a quark at the center of a large nucleus $\hat q_{N}$ has been studied in
deeply inelastic scattering (DIS) \cite{Wang:2002ri,Majumder:2004pt}.
A recently extracted value \cite{Deng:2009qb}  $\hat q_{N}\approx 0.02$ GeV$^{2}$/fm
from the HERMES \cite{Airapetian:2007vu} experimental data is used here.
All hadron resonances with mass below 1 GeV are considered for the calculation of the
hadron density at a given temperature $T$ and zero chemical potential.
A full 3+1D ideal hydrodynamics \cite{Hirano2001,HT2002} is used 
to provide the space-time evolution of the local temperature and flow velocity in the 
bulk medium along the jet propagation path in heavy-ion collisions. The initial highest temperatures $T_0$ in the
center of the most central heavy-ion collisions are set to reproduce the measured charged hadron
rapidity density. The initial spatial energy density distribution follows that of a Glauber model with Wood-Saxon nuclear
distribution.  At the initial time $\tau_0=0.6$ fm/$c$, $T_0=373$ and 473 MeV for Au+Au collisions at RHIC and
Pb+Pb collisions at LHC, respective.

\begin{figure}
\centering
\includegraphics[width=3.2in ]{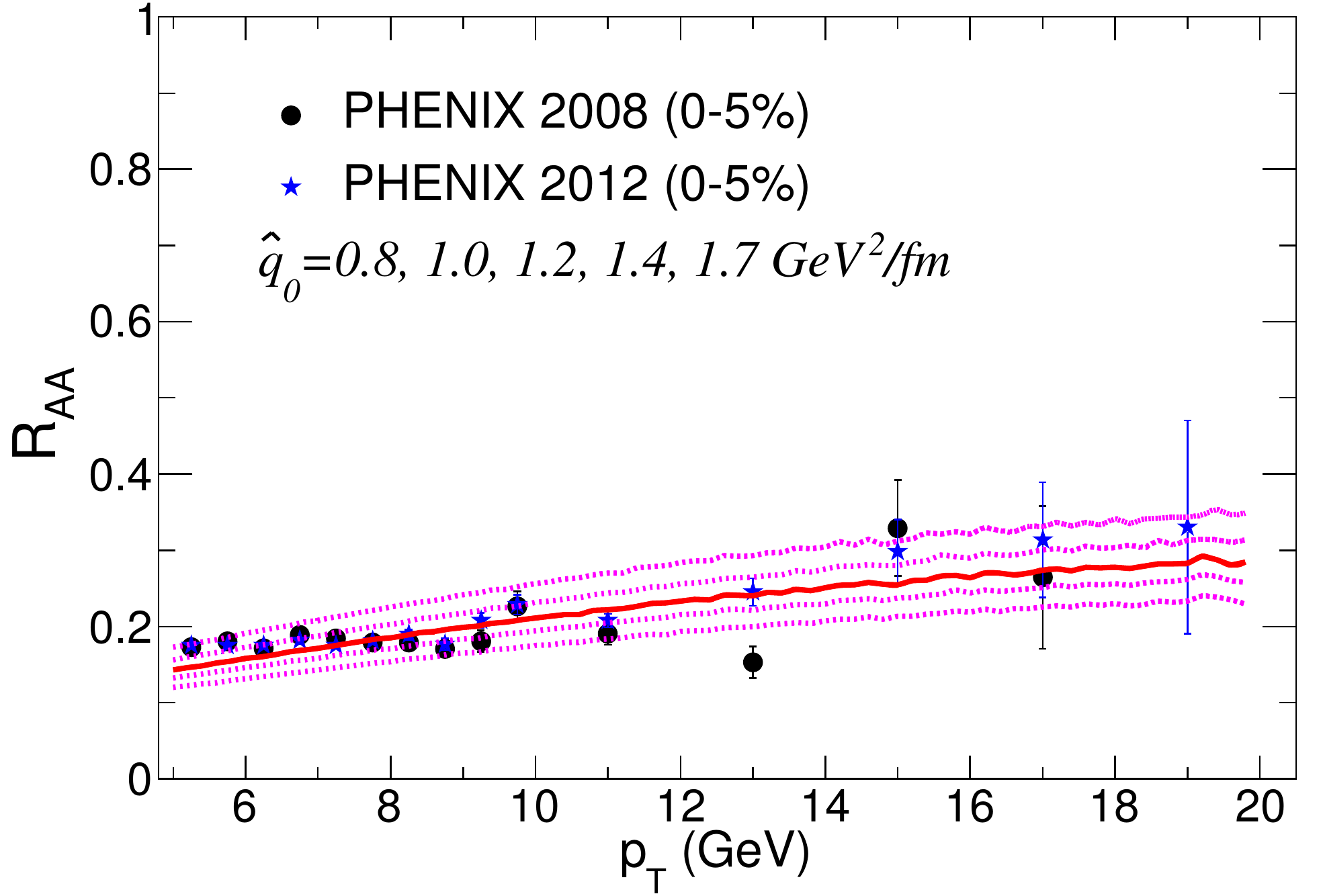}
\includegraphics[width=3.2in ]{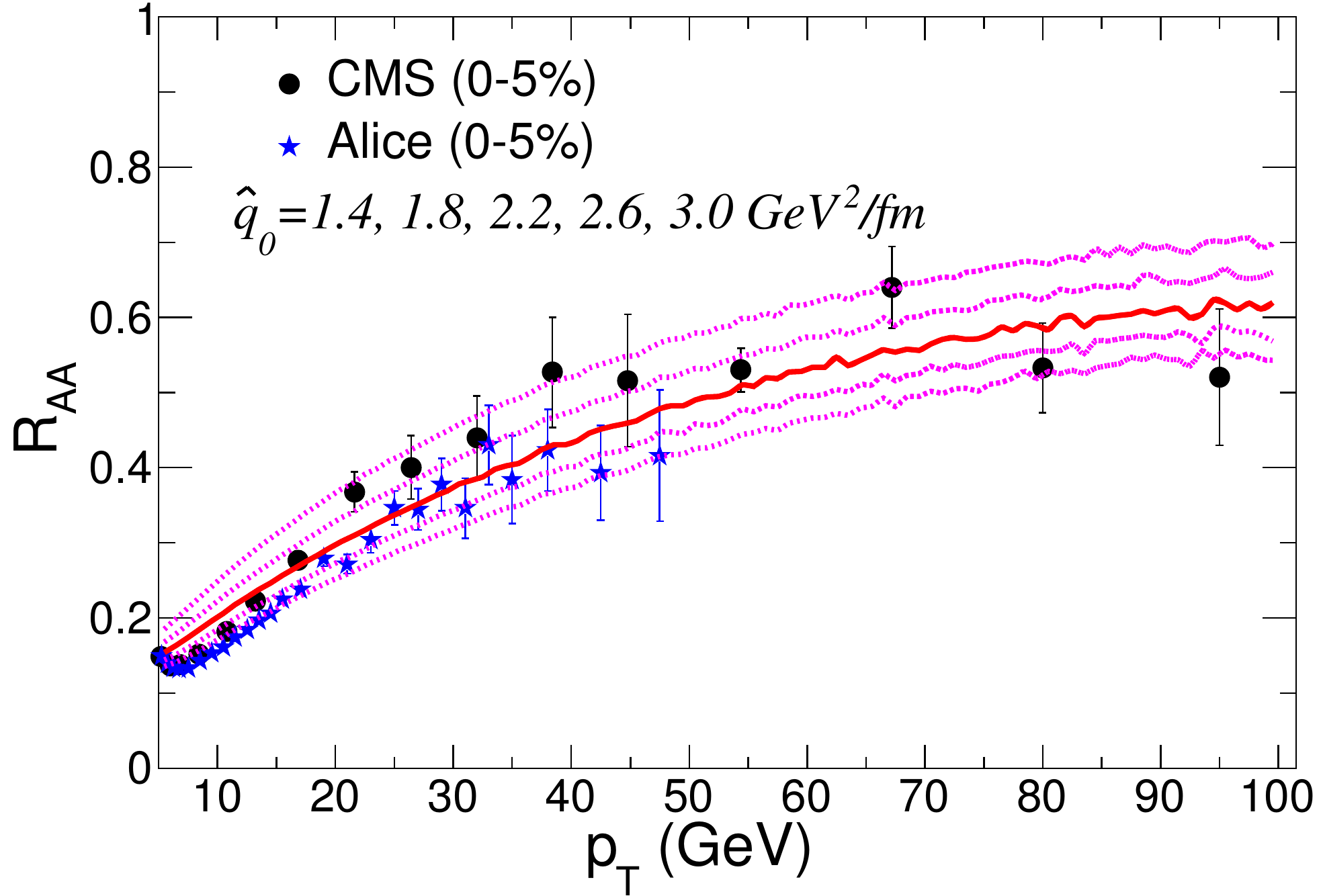}
\caption{(Color online) HT-BW results for the nuclear modification factor at mid-rapidity for neutral pion spectra in  $0-5\%$ central Au+Au collisions at $\sqrt{s}=200$ GeV/n (upper panel) and Pb+Pb collisions at $\sqrt{s}=2.76$ TeV/n (lower panel) 
with a range of values of  initial quark jet transport parameter $\hat q_{0}$ at $\tau_{0}=0.6$ fm/$c$ in the center of the most central collisions, as compared to PHENIX data \cite{Adare:2008qa,Adare:2012wg} at RHIC and ALICE \cite{Abelev:2012hxa} and CMS data \cite{CMS:2012aa} at LHC.} 
\label{fig:htbw-raa}
\end{figure}

\begin{figure}
\includegraphics[width=3.2in ]{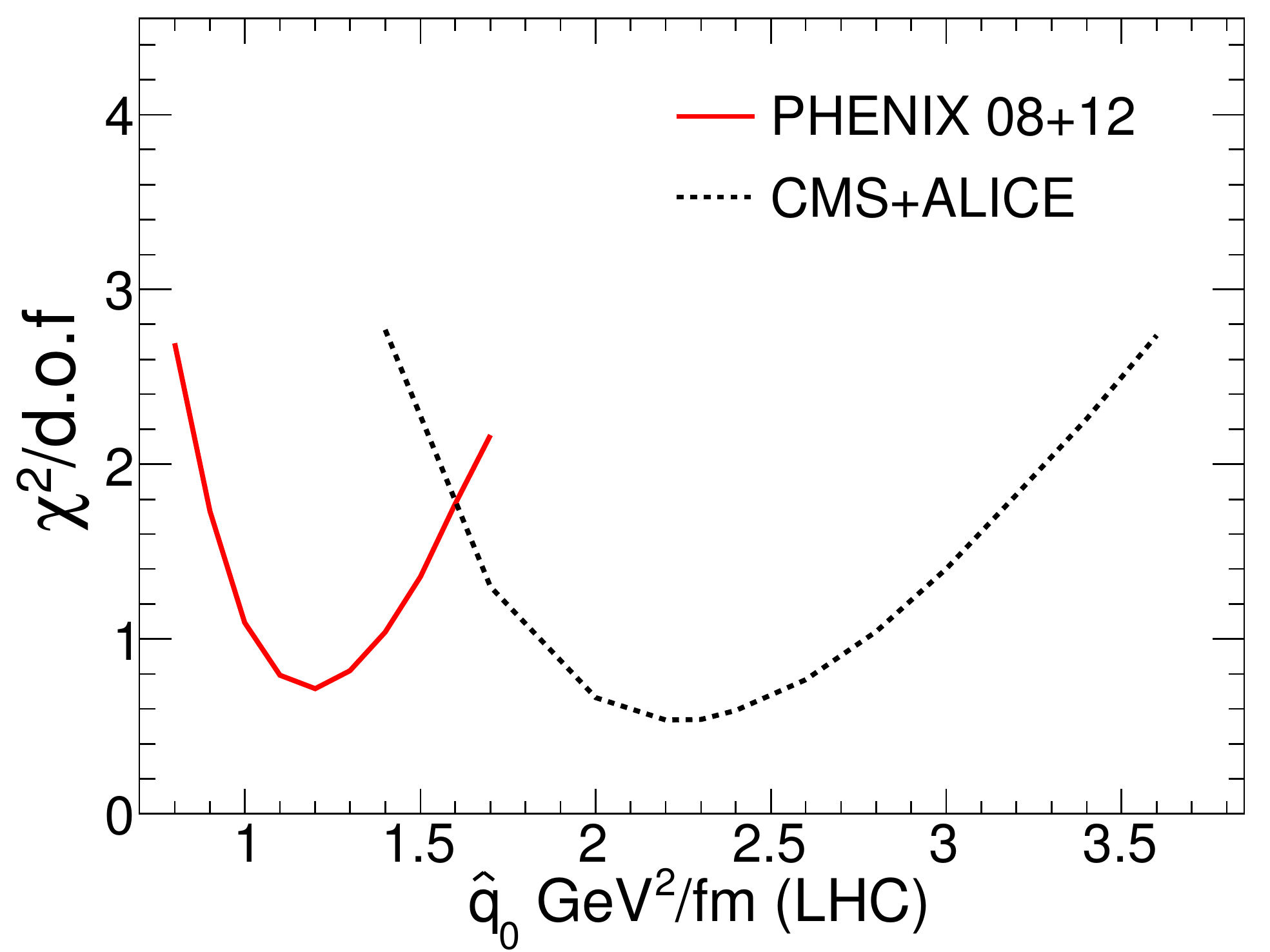}
\caption{ (Color online) The $\chi^2$/d.o.f  as function of the initial quark jet transport parameter $\hat q_0$ from fitting to the PHENIX data \cite{Adare:2008qa,Adare:2012wg}  (combined 2008 and 2012 data set) at RHIC for $p_T>5$ GeV/$c$ and combined ALICE \cite{Abelev:2012hxa} and CMS \cite{CMS:2012aa}  data at LHC for $p_T>15$ GeV/$c$ by the HT-BW model calculation of the nuclear suppression factor $R_{AA}(p_T)$ as shown in Fig.~\ref{fig:htbw-raa}.
\label{fig:htbw-chi2} } 
\end{figure}

With the above medium modified fragmentation functions and temperature dependence of the jet transport coefficient, one can calculate the nuclear modification factors and compare to the experimental data as shown in Fig.~\ref{fig:htbw-raa}. From $\chi^2$ fits to experimental data at RHIC and LHC  as shown in Fig.~\ref{fig:htbw-chi2}, one can extract values of quark jet transport parameter $\hat q_0$ at the center of the most central A+A collisions at a given initial time $\tau_0$. Best fits to the combined PHENIX data on neutral pion spectra \cite{Adare:2008qa,Adare:2012wg} in 0-5\% central $Au+Au$ collisions at $\sqrt{s}=0.2$ TeV/n gives
$\hat q_{0}=1.20 \pm 0.30 $ GeV$^2$/fm  (at $\tau_0=0.6$ fm/$c$).
Similarly, best fit to the combined ALICE \cite{Abelev:2012hxa} and CMS \cite{CMS:2012aa} data on changed hadron spectra in 0-5\% central Pb+Pb collisions at $\sqrt{s}=2.76$ TeV/n leads to $\hat q_{0}=2.2 \pm 0.5$  GeV$^2$/fm (at $\tau_0=0.6$ fm/$c$).

The charged hadron pseudo-rapidity density at mid-rapidity $dN_{ch}/d\eta=1584\pm 4 (stat.) \pm 76 (sys.)$
in the most central $0-5\%$ Pb+Pb collisions at $\sqrt{s}=2.76$ TeV/n as measured by ALICE
experiment \cite{Aamodt:2010pb} is $2.3\pm 0.24$ larger than $dN_{ch}/d\eta=687\pm 37$ for
 0-5\%  Au+Au collisions at $\sqrt{s}=0.2$ TeV/n \cite{phenix-nch}. Taking into account the difference in nuclear
 sizes, the ratio of the transverse hadron density in central Pb+Pb at LHC and Au+Au at RHIC is about $2.2\pm 0.23$.
 If one assumes that the jet transport coefficient  is proportional to the initial parton density or the transverse density of
charged hadron multiplicity in mid-rapidity, this should also be the ratio of the initial jet transport parameters in these
collisions at LHC and RHIC, which is very close to the value of $1.83\pm 0.26$ one obtains from independent fits 
to the experimental data at RHIC and LHC on hadron suppression factors.

\section{The Higher-Twist-Majumder (HT-M) model}
\label{sect:HT}

Similar to the HT-BW model, HT-M approach~\cite{Majumder:2010qh,Majumder:2009ge} is a straightforward evaluation of the 
first power correction to the vacuum evolution of a fragmentation function. It, however, goes beyond the single scattering and includes multiple induced gluon emission through a set of effective modified QCD evolution equations. 
One calculates  the medium modified fragmentation function by evolving an input fragmentation function using a vacuum plus medium modified 
kernel. As such, the formalism explicitly imbibes the concept of factorization~\cite{Collins:1985ue}: the initial parton distribution functions are 
factorized from the hard scattering cross section, these are also factorized from the final fragmentation function. The cross 
section to produce hadrons at a given transverse momentum $p_{h}$ and in a given rapidity interval $y$ may be expressed as, 
\begin{eqnarray}
\frac{d \sigma }{d y d^{2} p_{h}} &=& \int d^{2} b d^{2} r T_{AB} (b,r) \int dx_{a} dx_{b} \nonumber \\
&\times & G_{A}(x_{a}, Q^2)  G_{B}(x_{b},Q^{2}) \frac{d \hat{\sigma}}{d \hat{t}} \frac{\tilde{D}(z , Q^2)}{\pi z} . \label{diff-cross}
\end{eqnarray}
In the equation above $T_{AB} (b,r) = \int dz \rho_{A}(z,\vec{r}+\vec{b}/2) \int dz' \rho_{B}(z',\vec{r}-\vec{b}/2) $, where $\rho_{A/B}$ represents the 
nuclear density in nucleus $A/B$. The nuclear parton distribution functions $G_{A} (x_{A},Q^{2})$ and $G_{B} (x_{B},Q^{2})$ are inclusive of any 
shadowing corrections. 
The modified fragmentation function $\tilde{D}$ contains two contributions: one from vacuum evolution which is contained in the regular DGLAP equations:
\begin{equation}
\frac{\partial D_q^h (z, Q^2)}{ \partial \log(Q^2) } = \frac{\alpha_S(Q^2)}{2\pi} \int_z^1 \frac{dy}{y} 
P_{q \to i} (y) D_i^h \left( \frac{z}{y}, Q^2 \right) . 
\label{HT:vac_DGLAP}
\end{equation}

The second contribution to the modified fragmentation function is from the medium modified evolution equation~\cite{Majumder:2009zu}, 
\begin{eqnarray}
\frac{\partial {D_q^h}(z,Q^2\!\!,q^-)|_{\zeta_i}^{\zeta_f}}{\partial \log(Q^2)}  &=& \frac{\alpha_S}{2\pi} \int\limits_z^1 \frac{dy}{y} 
\int\limits_{\zeta_i}^{\zeta_f} d\zeta {P}(y) K_{q^-,Q^2} ( y,\zeta) \nonumber \\
&\times& {D_q^h}\left. \left(\frac{z}{y},Q^2\!\!,q^-y\right) \right|_{\zeta}^{\zeta_f}.  \label{HT:in_medium_evol_eqn}
\end{eqnarray}
In both Eqs.~\eqref{HT:vac_DGLAP} and \eqref{HT:in_medium_evol_eqn}, the splitting function $P_{q \to i} (y)$ is the regular 
Altarelli-Parisi splitting function. The modification from the medium is contained in the factor $K_{q^-,Q^2} ( y,\zeta) $. 
All factors of the medium (such as the transport coefficients $\hat{q}$) are contained within this factor, along with phase 
factors that arise due to interference between different amplitudes of emission. The contribution to $K$ from the leading power 
correction is given as, 
\begin{eqnarray}
K_{q^-, Q^2} ( y, \zeta ) &=& \frac{ \left[ \hat{q}_A(\zeta) -  (1-y) \hat{q}_A/2 + (1-y)^2 \hat{q}_F  \right]}{Q^{2}}
\nonumber \\
&\times& \left[ 2 - 2 \cos\left( \frac{Q^2 (\zeta - \zeta_i)}{ 2 q^- y (1- y)} \right)  \right] .
\end{eqnarray}
In the equation above, $\zeta$ and $\zeta_{i}$ represent the location of scattering and location of origin of the hard 
parton, respectively. The position ($\zeta$) dependent jet transport coefficient of a
gluon, $\hat{q}_A(\zeta)$, can be expressed in operator form \cite{Majumder:2007hx,Idilbi:2008vm,Majumder:2012sh},
similarly as in Eq.~(\ref{eq:qhat-ff}) except the color factor for a gluon jet $C_F\rightarrow C_A$.
Note that the $\hat{q}$ for a quark scattering off the gluon field is trivially related to the above expression as 
$\hat{q}_F= \frac{C_{F}}{C_{A}} \hat{q}_A $. 

\begin{figure}[htbp]
\resizebox{3.0in}{2.2in}{\includegraphics{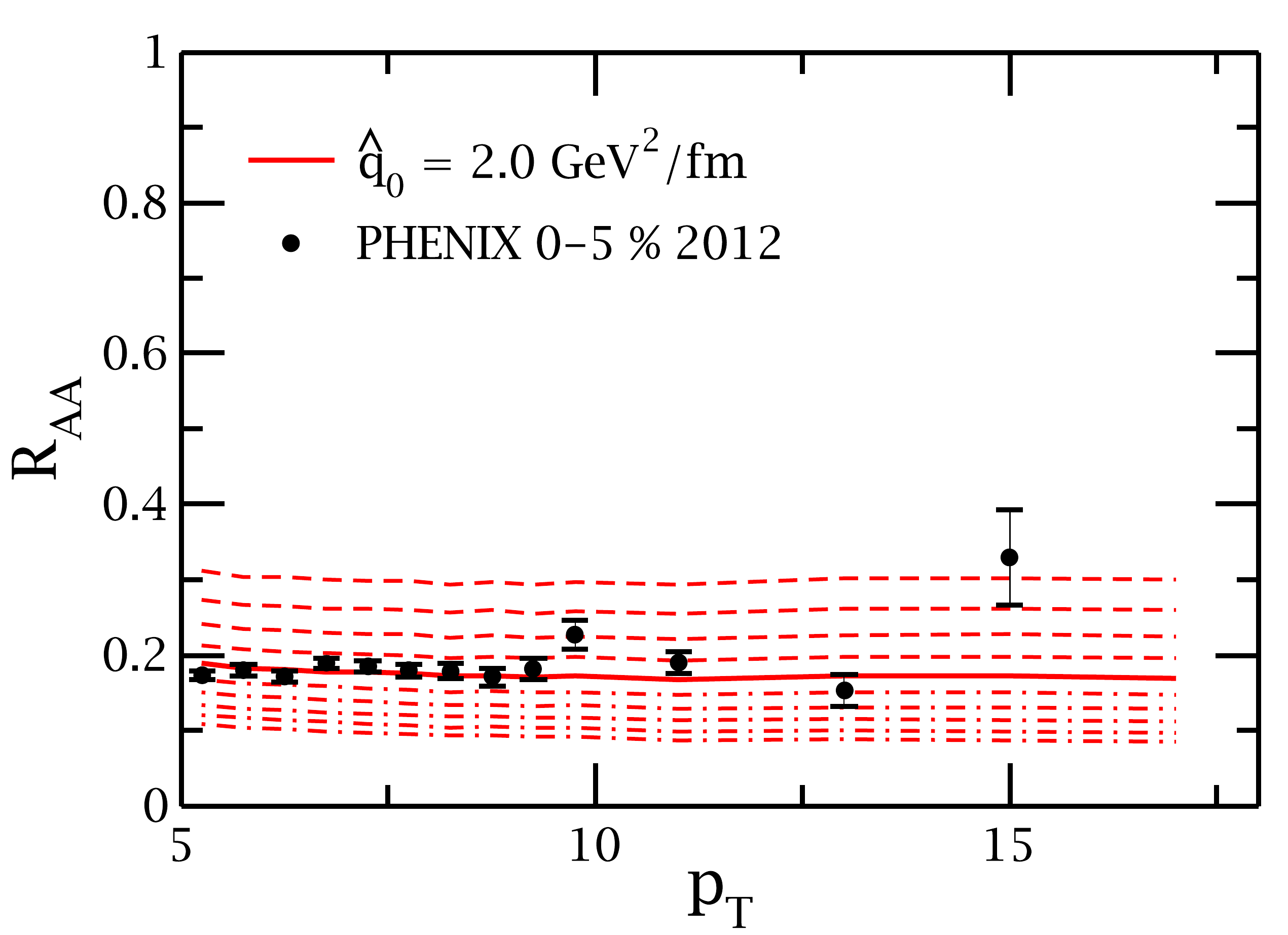}} 
\resizebox{3.0in}{2.2in}{\includegraphics{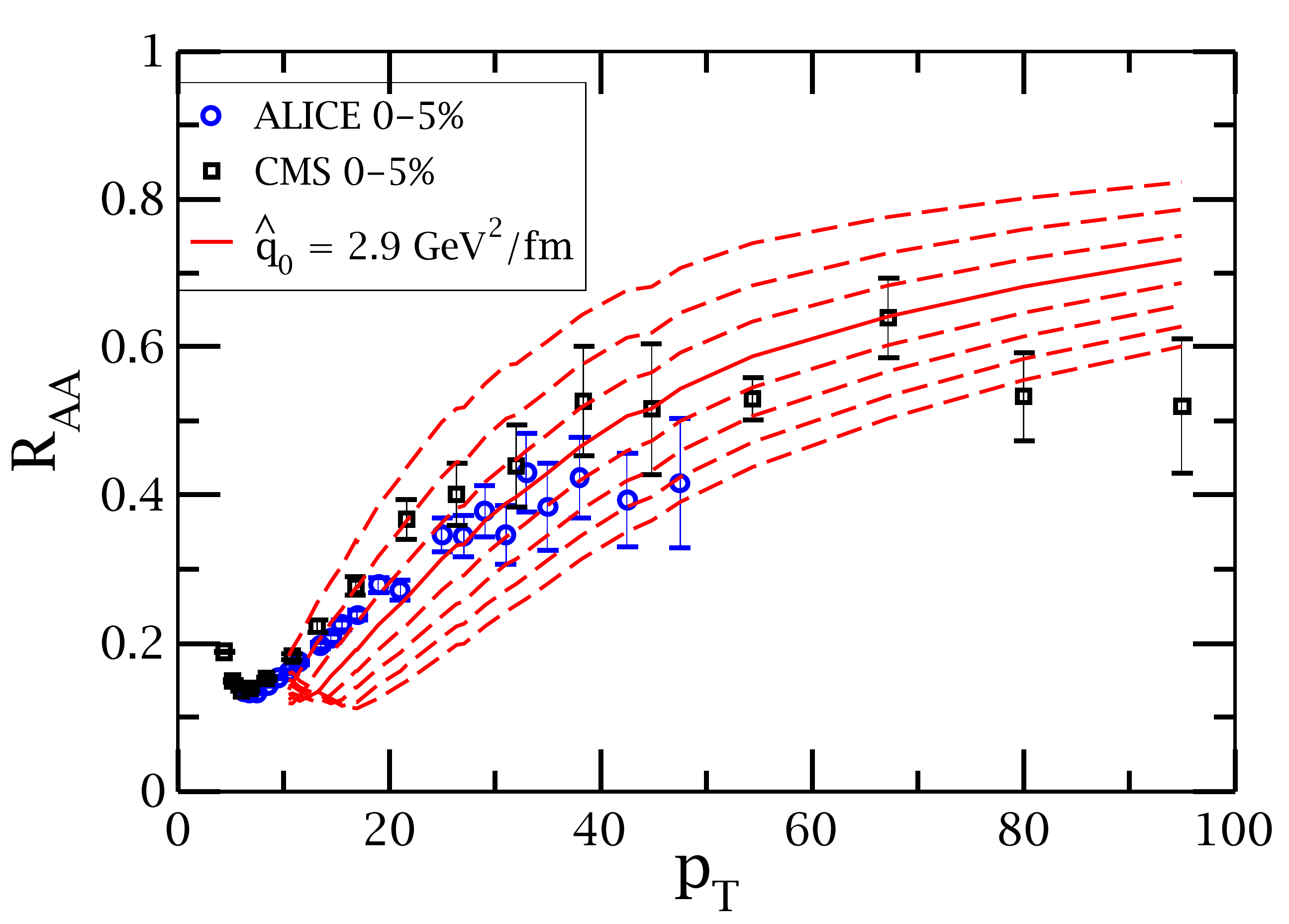}} 
\caption{(Color online) HT-M results for the nuclear modification factor at mid-rapidity for neutral pion spectra in  $0-5\%$ central Au+Au collisions at $\sqrt{s}=200$ GeV/n (upper panel) and Pb+Pb collisions at $\sqrt{s}=2.76$ TeV/n (lower panel) 
with a range of values of  initial gluon jet transport parameter $\hat q_{0}$ (at $\tau_{0}$=0.6 fm/$c$) in the center of the most central collisions, as compared to PHENIX data \cite{Adare:2008qa,Adare:2012wg} at RHIC and ALICE \cite{Abelev:2012hxa} and CMS data \cite{CMS:2012aa} at LHC.} 

\label{fig:htm-raa}
\end{figure}

In actual calculations of the nuclear modification factor, one assumes $\hat{q}$ to scale with some intrinsic quantity 
in the medium. In the calculations presented in this section,  $\hat{q}$ is assumed to scale with the entropy density $s$ (see Refs.\cite{Bass:2008rv,Majumder:2007ae} for other scalings assumptions for $\hat q$):
\begin{equation}
\hat{q}(s) = \hat{q}_{0} \frac{s}{s_{0}}.
\end{equation}
In the equation above, $s_{0}$ is the maximum entropy density achieved at an initial time $\tau_0$ 
in the center of the most central collisions 
at top RHIC energy. The value of $\hat{q} = \hat{q}_{0}$ corresponds to this point. The space-time evolution of the 
entropy density is given by (2+1)D viscous hydrodynamic model  \cite{Shen:2010uy,Shen:2011eg} tabulated by the hydro group within the JET Collaboration. These hydro profiles are obtained with MC-KLN initial conditions in which the initial temperature is $T_0=346$ MeV at the center of the most central Au+Au collisions at RHIC ($\sqrt{s}=200$ GeV/n)  and 447 MeV in Pb+Pb collisions at LHC ($\sqrt{s}=2.76$ TeV/n). In the calculation of the hadron spectra in heavy-ion collisions, the distance integral over $K$ is then sampled over a large number of paths passing through the evolving medium. The starting points of all the paths are obtained by sampling the binary collision profile. The medium averaged length integral over $K$ is then used to calculate the medium modified evolution of the fragmentation function using Eqs.~\eqref{HT:vac_DGLAP} and \eqref{HT:in_medium_evol_eqn}. 

\begin{figure}
\includegraphics[width=3.0in ]{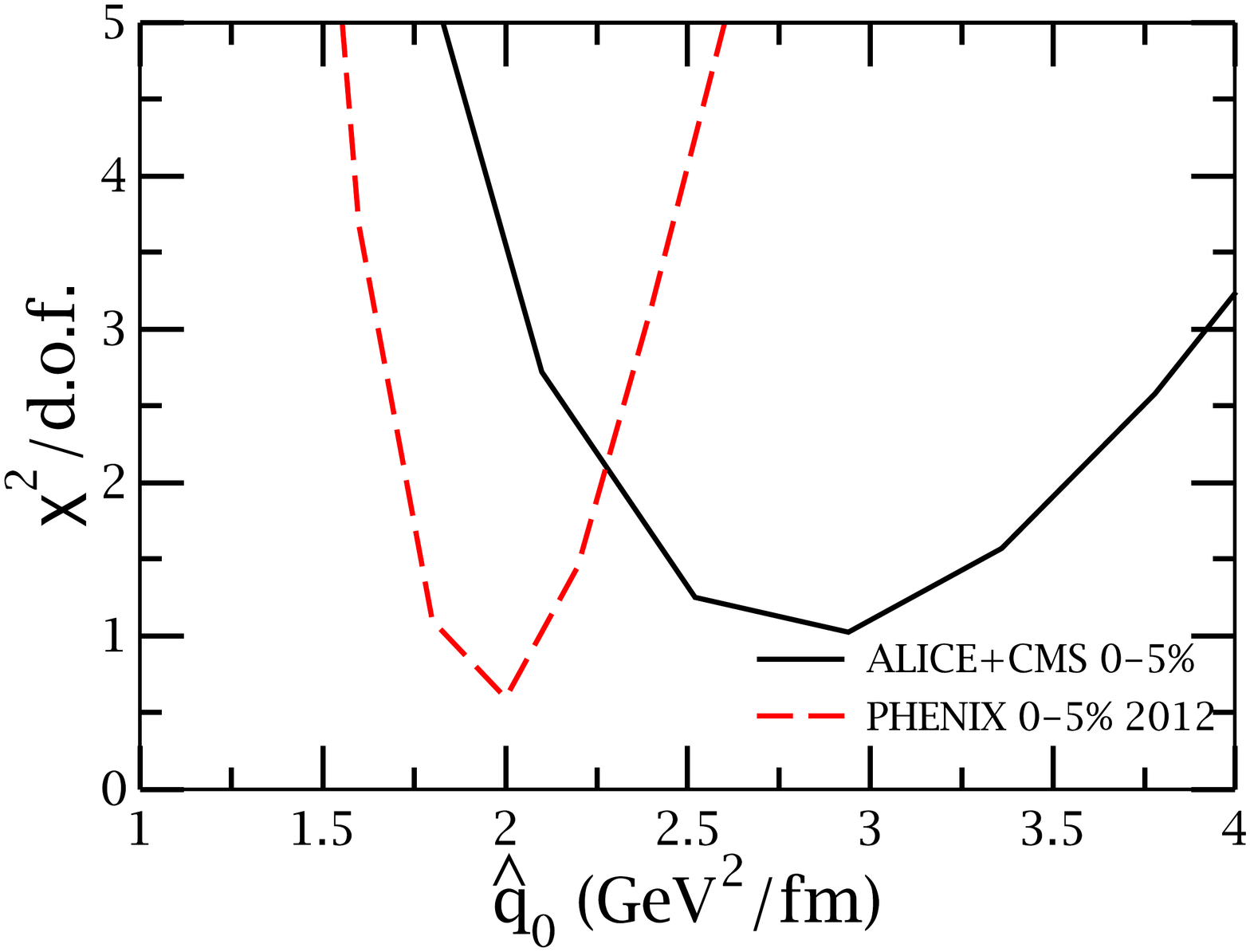}
\caption{ (Color online) The $\chi^2$/d.o.f  as function of the initial gluon jet transport parameter $\hat q_0$ from fitting to the PHENIX data \cite{Adare:2008qa,Adare:2012wg}  (combined 2008 and 2012 data set) at RHIC and combined ALICE \cite{Abelev:2012hxa} and CMS \cite{CMS:2012aa}  data at LHC by the HT-M model
calculation of the nuclear suppression factor $R_{AA}(p_T)$ as shown in Fig.~\ref{fig:htm-raa}.}
\label{fig:htm-chi2}
\end{figure}

Both medium and vacuum evolution equations require an input distribution. This is taken as a vacuum fragmentation function at the 
input scale of $Q_{0}^2 = p/L$, where $p = p_{h} / z$ is the transverse momentum of the parton which fragments 
to a hadron with transverse momentum $p_{h}$ with a momentum fraction $z$.  Such input vacuum fragmentation functions are evolved
according to the vacuum evolution equations from $Q_0^2=1$ GeV$^2$. 
The factor $L$ is the mean escape length of jets of that energy in the medium. The mean escape length is calculated 
by calculating the maximum length that could be travelled by a parton with an energy $p$ using the single emission formalism of Guo and 
Wang~\cite{Guo:2000nz,Wang:2001ifa}.

The results presented in the following represent updates of calculations that have appeared in Ref.~\cite{Majumder:2011uk}. The fluid dynamical simulations have be been updated to include a new initial state and averaged over an ensemble of fluctuating initial conditions \cite{Qiu:2011hf,Qiu:2012uy}. Unlike previous calculations, the binary collision profile which determines the distribution of jet origins is also consistently determined by averaging over the same ensemble of initial conditions.

In Fig.~\ref{fig:htm-raa}, calculations of the hadron suppression factor in $0-5\%$ central Au+Au collisions at RHIC ($\sqrt{s}=200$ GeV/n) (upper panel) and  $0-5\%$ central Pb+Pb collisions at LHC ($\sqrt{s}=2.76$ TeV/n) (lower panel) are compared to the experimental data.
The lines represent calculated values of $R_{AA}$ for different values of initial values of $\hat q_0$ at the center of of most central heavy-ion collisions.
The solid lines represent the best fit to the experimental data.  The range of $p_T$ of the fits are $p_T\ge 5$ and 20 GeV/$c$ at RHIC and LHC, respectively. Shown in Fig.~\ref{fig:htm-chi2} are the $\chi^2$ distributions as a function of the initial value of $\hat q_0$ from fits to the experimental data as in Fig. ~\ref{fig:htm-raa}. The values of the jet transport parameter from the 
best fits are $\hat q_0=2.0 \pm 0.25 $ GeV$^2$/fm and $2.9 \pm 0.6$ GeV$^2$/fm at RHIC and LHC, respectively.

\section{MARTINI Model}

In the factorized picture,
jet production in relativistic heavy ion collisions proceeds in stages.
The first stage is the collision of initial state partons. Since the energy
and the virtuality of these partons are $O(\sqrt{s})$, this stage takes
place well before the formation of QGP. The second stage is the propagation
of the scattered partons in the produced QGP. In the MARTINI approach of jet quenching 
\cite{Qin:2007rn,Schenke:2010nt,Schenke:2009gb}, 
the nuclear initial parton scatterings for jet production
are carried out by using PYTHIA-8 on
each nucleon-nucleon collision with Glauber geometry.
The propagation of jet partons is then 
carried out by solving the following rate equations using Monte-Carlo
methods
\begin{widetext}
\begin{eqnarray}
 \frac{d\Pg (p)}{dt} &=&  \int_k \!\!
        \Pq (p{+}k) \frac{d\Gamma^q_{\!qg}(p{+}k,p)}{dk}  
        {+}\int_k\!\!\Pg (p{+}k)\frac{d\Gamma^g_{\!\!gg}(p{+}k,k)}{dk}
        -\int_k\!\!\Pg (p) \left[\frac{d\Gamma^g_{\!q \bar q}(p,k)}{dk}
        + \frac{d\Gamma^g_{\!\!gg}(p,k)}{dk} \Theta(k{-}p/2)\right]
     \, , \nonumber \\
\frac{d\Pq (p)}{dt} &=&  \int_k \!\!
      \Pq (p{+}k) \frac{d\Gamma^q_{\!qg}(p{+}k,k)}{dk}  
 -\int_k\!\!\Pq (p)\frac{d\Gamma^q_{\!qg}(p,k)}{dk}
 +2 \int_k\!\!\Pg (p{+}k)\frac{d\Gamma^g_{\!q \bar q}(p{+}k,k)}{dk}
\end{eqnarray}
\end{widetext}
where $d\Gamma^a_{bc}(p,k)/dk$ is 
the $a\to b+c$ splitting rate 
calculated in the full leading order thermal QCD
that includes the HTL effects and the LPM effects. 
All split partons with energy above a threshold (currently set to 4 times
the local temperature) are tracked of until the partons fragment outside
of QGP.  Elastic scatterings are included in a similar way.

In this approach, the properties of the local medium \cite{Nonaka:2006yn,Schenke:2010nt,Schenke:2010rr} enter through the
local temperature and the flow velocity when calculating the rates, and the
interaction between the parton and the medium is controlled by the HTL
resummed elastic collision rate
\be
{d\Gamma_{\rm el}\over d^2\bfqperp}
= {C_a\over (2\pi)^2} {g^2 m_D^2 T\over \bfqperp^2(\bfqperp^2 + m_D^2)}
\ee
where $T$ is the fluid rest frame temperature,
 $g$ is the coupling constant of the strong interaction and 
$m_D^2 = g^2 T^2(2N_c + N_f)/6$ is the Debye mass squared. The factor
$C_a$ is the Casimir of the propagating parton.
Hence, the average transverse momentum transfer squared per mean free path,
$\hatq = {\ave{\bfqperp^2}/l_{\rm mfp}}$, is not a primary parameter
of the calculation but a derived quantity.
In the fluid rest frame, it is given by
\be
\hatq = \int^{q_{\rm max}} d^2\bfqperp\, \bfqperp^2 
{d\Gamma_{\rm el}\over d^2\bfqperp}
\ee
where $q_{\rm max}$ is the UV cut-off. In a static
medium, it is given by
\be
\hatq
= C_a {\alpha_{\rm s} m_D^2 T \ln(1 + q_{\rm max}^2/m_D^2)}
\label{eq:localqhat}
\ee
where $q_{\rm max}\approx 6ET$.

This model can describe the suppression of hadron spectra in heavy-ion collisions at RHIC very well with a fixed value of strong coupling constant \cite{Schenke:2009gb}. 
For the LHC $R_{AA}$ calculation, it is necessary to include a running coupling
constant in the splitting kernel since the kinematic range at the LHC is much
wider than that at RHIC. Currently, the coupling constant for the elastic
scattering is treated as a constant, however there is no technical difficulty to
include the scale dependence. The vertex momentum scale is determined by 
$\langle |k_T| \rangle = (\hat{q}k)^{1/4}$. The rates also includeÊfinite-size 
effects via a parametrization of the rates derived in \cite{CaronHuot:2010bp}.

\begin{figure}[t]
\centerline{
\includegraphics[width=3.5in]{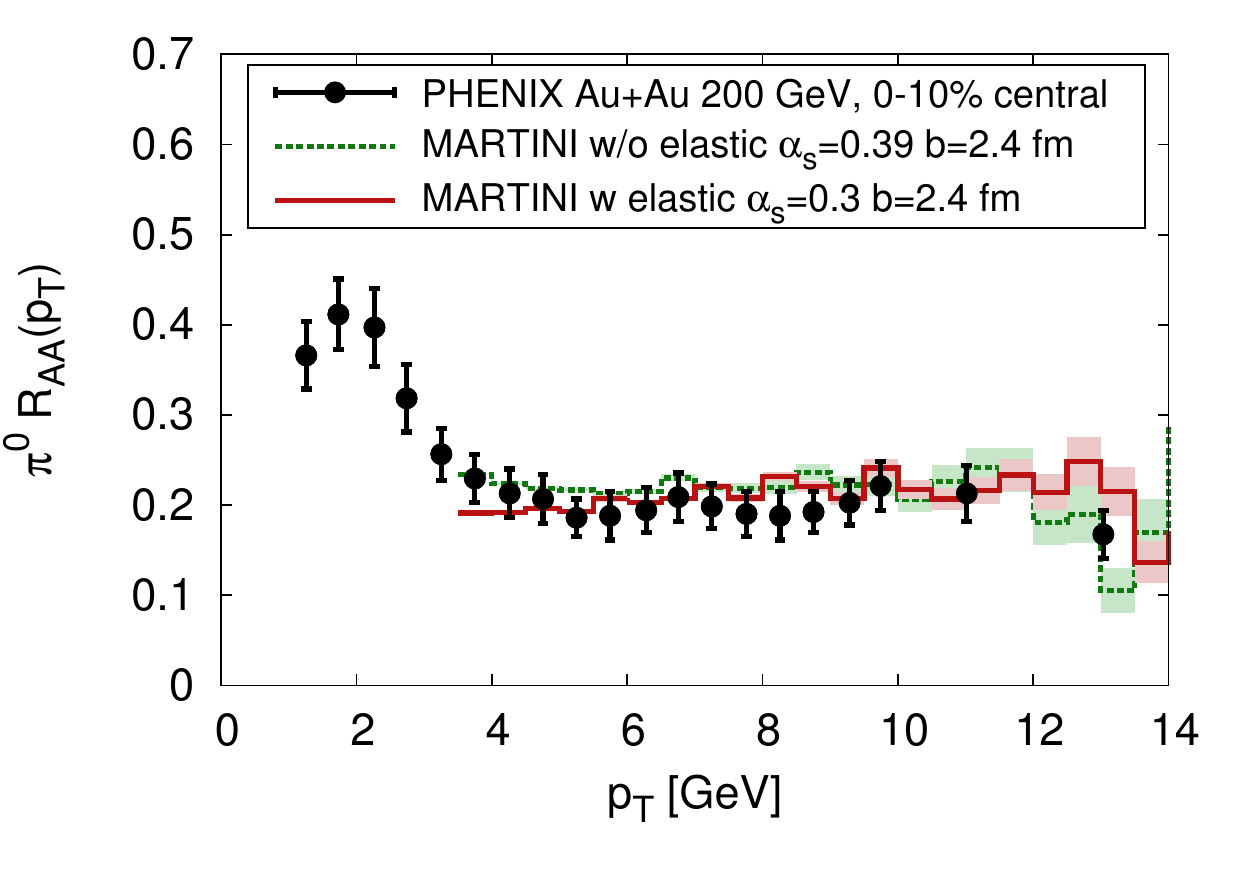}}
\vspace{-12pt}
\centerline{
\includegraphics[width=3.3in]{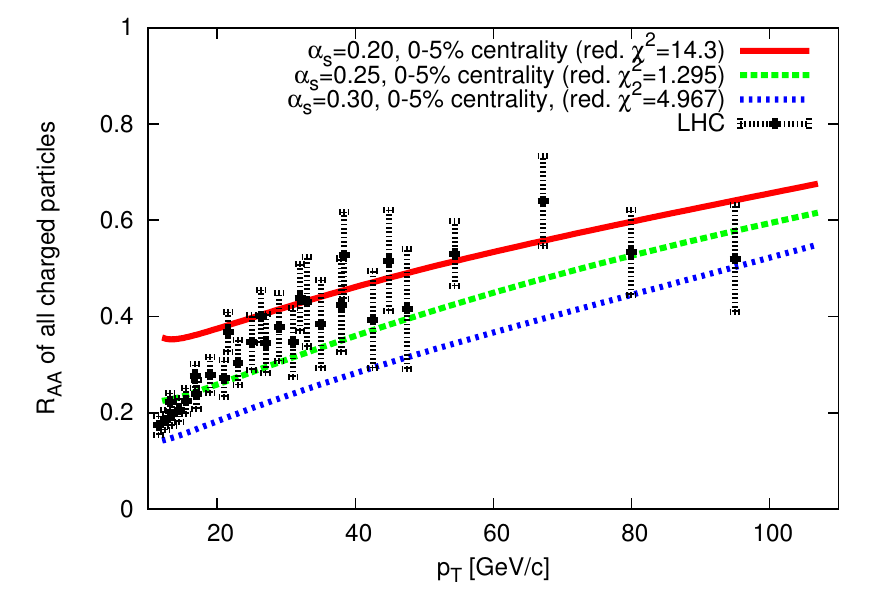}}
\caption{(Color online) The nuclear modification factors for the central Au+Au collisions at RHIC (upper panel) (from Ref.~\cite{Schenke:2009gb})  and Pb-Pb collisions at
the LHC (lower panel) from the MARTINI model as compared to
PHENIX data \cite{Adare:2008qa,Adare:2012wg}  at RHIC and combined 
ALICE \cite{Abelev:2012hxa} and CMS \cite{CMS:2012aa} data at LHC.
The bulk medium space-time profile is given by (3+1)-dimensional simulations by Nonaka and Bass \cite{Nonaka:2006yn}
for Au+Au collisions at RHIC and  by the MUSIC 3+1D ideal 
hydrodynamics calculation \cite{Schenke:2010nt} for Pb+Pb at LHC.
}
\label{fig:amy-raa}
\end{figure}

Shown in Fig.~\ref{fig:amy-raa} are the calculated single hadron suppression factor $R_{AA}(p_T)$ in central Au+Au
collisions at RHIC from Ref.~\cite{Schenke:2009gb} and Pb+Pb collisions at LHC as compared to the experimental data. 
For Au+Au collisions at RHIC, the best fit to the data within MARTINI is achieved with $\alpha_{\rm s} = 0.30$. 
For LHC, the best fit is achieved with $\alpha_{\rm s} = 0.25$. With these values of strong coupling constant, one can calculate both the temperature and energy dependence of the jet transport coefficient according to Eq.~(\ref{eq:localqhat}).

\section{McGill-AMY model}

Similar to MARTINI model, the basic scattering and radiation processes in McGill-AMY model are also described by the
thermal QCD with HTL effects and LPM interference. However, instead of a full Monte Carlo approach, McGill-AMY employs
collinearized interaction rates (integrated over the transverse momentum) and numerically solve the rate equations for parton momentum distributions. These distributions are then convoluted with pQCD cross sections for initial jet production and parton fragmentation functions to give the final hadron spectra in heavy-ion collisions.

In this approach, the evolution of hard jets (quarks and gluons) in the hot QCD medium is obtained by solving a set of rate equations for their momentum distributions $f(E,t) = {dN(p,t)}/{dp}$. The generic form of these rate equations may be written as:
\begin{eqnarray}
\frac{df_j(p,t)}{dt} \!&=&\! \sum_{ab} \! \int \! dk \left[f_a(p+k,t) \frac{d\Gamma_{a\to
j}(p+k,k)}{dk dt} \right. \nonumber \\
&-& \left. P_j(k,t)\frac{d\Gamma_{j\to b}(p,k)}{dkdt}\right],
\end{eqnarray}
where $d{\Gamma_{j\to a}(p,k)}/{dk dt}$ is the transition rate for the partonic process $j\to a$, with $p$
the initial jet energy and $k$ the energy lost in the process. The energy gain channels are taken into account by the integration for the $k<0$ part. The radiative parts of the transition rates are taken from Ref. \cite{Arnold:2002ja, Jeon:2003gi}; for the collisional parts, the contributions from the drag and the diffusion are included as in Ref. \cite{Qin:2007rn, Qin:2009bk}.

After solving the above rate equations, the medium-modified fragmentation function
$\tilde{D}_{h/j}(z,\vec{r}_\bot, \phi)$ for a single partonic jet may be obtained as follows,
\begin{eqnarray}
\label{mmff} \tilde{D}_{h/j}(z,\vec{r}_\bot, \phi) \!&=&\! \sum_{j'} \!\int\! dp_{j'} \frac{z'}{z}
D_{h/j'}(z') P(p_{j'}|p_j,\vec{r}_\bot, \phi), \nonumber \\
\end{eqnarray}
where $z = p_h / p_{j}$ and $z' = p_h / p_{j'}$ are two momentum fractions, with $p_h$ the hadron momentum and $p_{j}$($p_{j'}$) the initial (final) jet momentum; $D_{h/j}(z)$ is the vacuum fragmentation function.
$P(p_{j'}|p_j,\vec{r}_\bot, \phi)$ represents the probability of obtaining a jet $j'$ with momentum $p_{j'}$ from a given jet $j$ with momentum $p_j$. It depends on the path taken by the parton and the medium
profiles (such as the temperature and flow velocity) along that path, which in turn depend on the production location $\vec{r}_\bot$ of the jet, and on its propagation direction $\phi$.
For the space-time evolution profiles (energy/entropy density, temperature and flow velocities) of the bulk QGP medium that jets interact with, we employ a (2+1)D viscous hydrodynamics model (VISH2+1) developed by The Ohio State University group \cite{Song:2007fn, Song:2007ux,Qiu:2011hf,Qiu:2012uy}, with two-component Glauber model for hydrodynamics initial conditions. The code version and parameter tunings for Pb+Pb collisions at LHC energies are taken as in Ref. \cite{Qiu:2011hf,Qiu:2012uy}. The highest initial temperature in the most central Au+Au collisions at RHIC and Pb+Pb collisions at LHC are $T_0=378$ MeV and 486 MeV, respectively, at an initial time $\tau_0=0.6$ fm/$c$.  When the local temperature of the medium drops below the transition temperature of $160$ MeV, jets are decoupled from the medium.

By convoluting the medium modified fragmentation function with the initial parton momentum distribution as computed from perturbative QCD calculations, one may obtain the hadron spectra:
\begin{eqnarray}
\frac{d\sigma_{AB\to hX}}{d^2p_T^hdy} \!&=&\! \int d^2\vec{r}_\perp {\cal
P}_{AB}(\vec{r}_\perp)  \sum_{j} \int \frac{dz}{z^2} \nonumber \\
&\times& \tilde{D}_{h/j}(z,\vec{r}_\bot, \phi) \frac{d\sigma_{AB\to jX}}{d^2p_T^j dy}.
\end{eqnarray}
The above equation contains the average over transverse positions $\vec{r}_\perp$ of initial hard jets via the probability distribution function ${\cal P}_{AB}(b,\vec{r}_\perp)$, which is determined from Glauber model simulation of binary collision distribution.  The propagation direction $\phi$ may be fixed or averaged over a certain range.
Putting all the ingredients together, one obtains the total yield of hadrons produced in relativistic nuclear collisions, which are used to calculate the nuclear modification factor $R_{AA}$.

\begin{figure}[htb]
\includegraphics[width=3.2in]{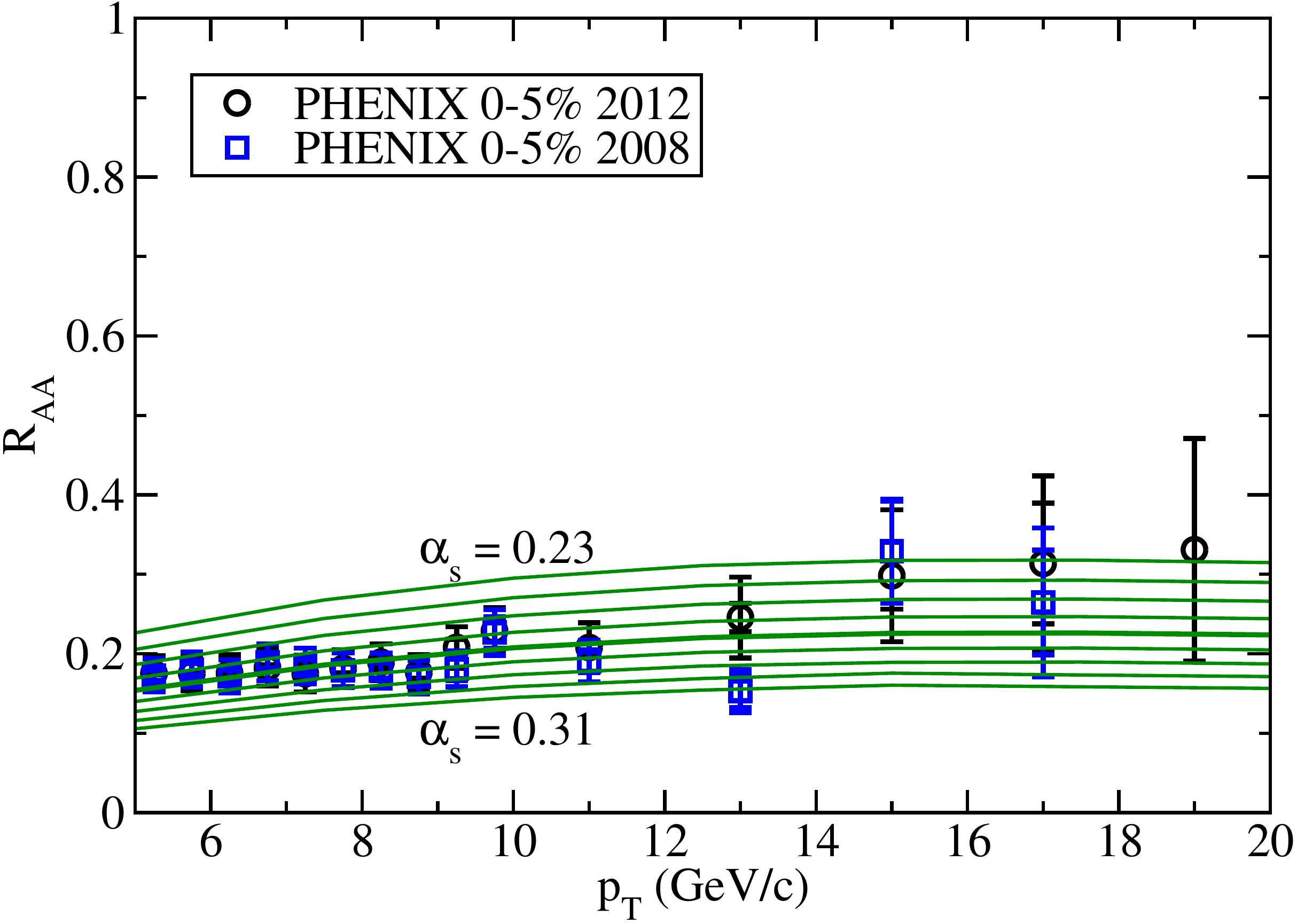}
\includegraphics[width=3.2in]{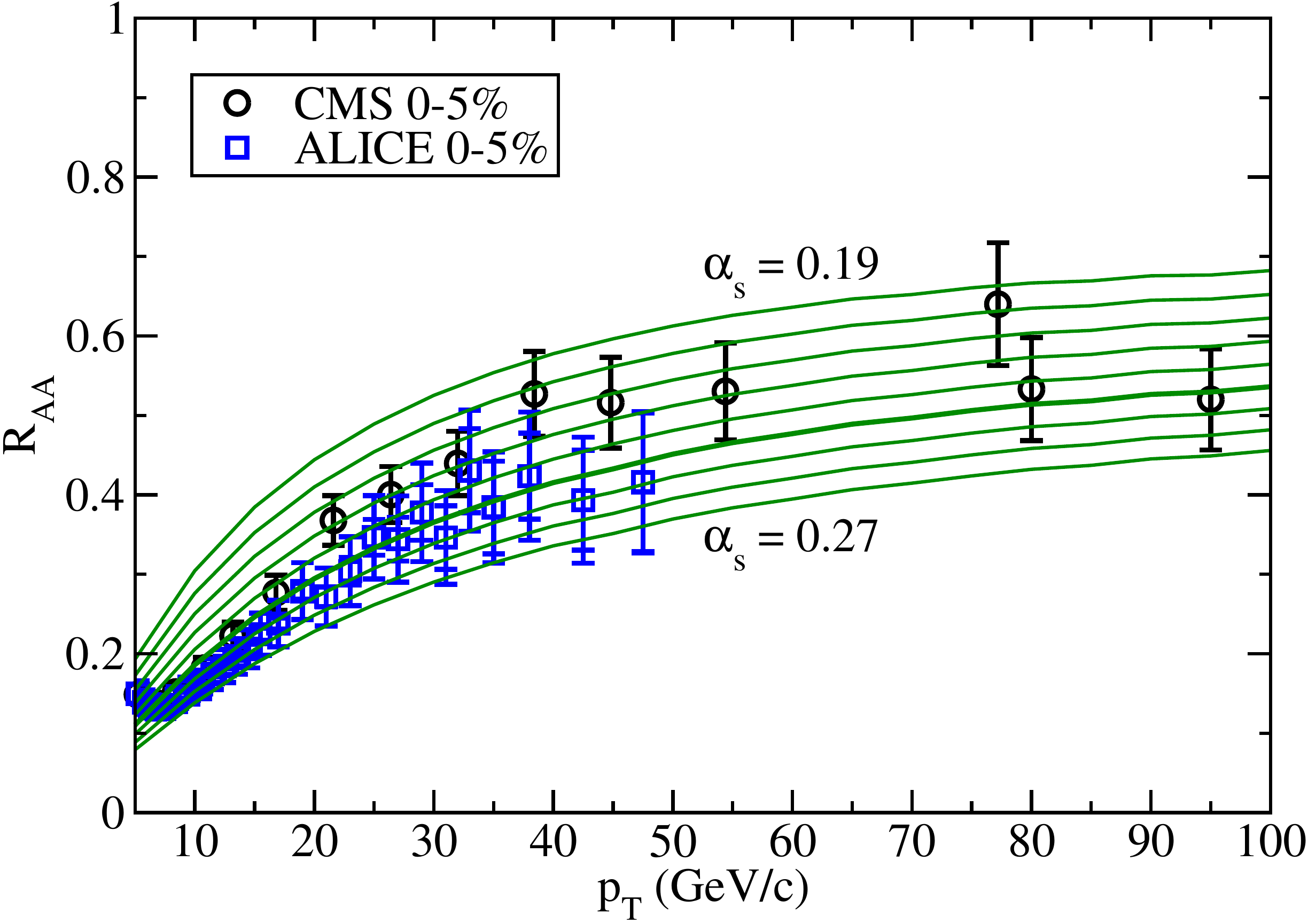}
\caption{(Color online) The nuclear modification factors $R_{AA}$ from McGill-AMY model as a function of $p_T$ for 0-5\% Au+Au collisions at RHIC (lower panel)
and 0-5\% Pb+Pb collisions at the LHC. Experimental data are taken from PHENIX experiment \cite{Adare:2008qa,Adare:2012wg} at RHIC and CMS  \cite{CMS:2012aa} and ALICE experiment \cite{Abelev:2012hxa} at LHC.  For difference curves from the top to the bottom, the values of $\alpha_s$ are from 0.23 to 0.31 with an increment of 0.1.} 
\label{mcgill-raa}
\end{figure}

In the upper panel of Fig. \ref{mcgill-raa},  the calculated suppression factors $R_{AA}$ for central 0-5\% collisions at RHIC
for different values of the fixed coupling constant $\alpha_s$ varies from $0.23$ to $0.31$ from the top to the bottom, with an increment of $0.1$, are compared to the experimental measurements taken from PHENIX Collaboration \cite{Adare:2008qa,Adare:2012wg}. The best fit to the experimental data is the thick curve in the middle, with $\alpha_s = 0.27 (+0.02/-0.015)$.

The lower panel of Fig. \ref{mcgill-raa} shows the comparison between the calculated $R_{AA}$ for central 0-5\% collisions at the LHC and experimental measurements taken from CMS \cite{CMS:2012aa} and ALICE Collaborations \cite{Abelev:2012hxa}. Calculations for different values of the fixed coupling constant $\alpha_s$ varies from $0.19$ to $0.27$ from the top to the bottom, with an increment of $0.1$. The best fit to the experimental data is the thick curve in the middle, with $\alpha_s = 0.24 (+0.02/-0.01)$.

\begin{figure}[htb]
\includegraphics[width=3.2in]{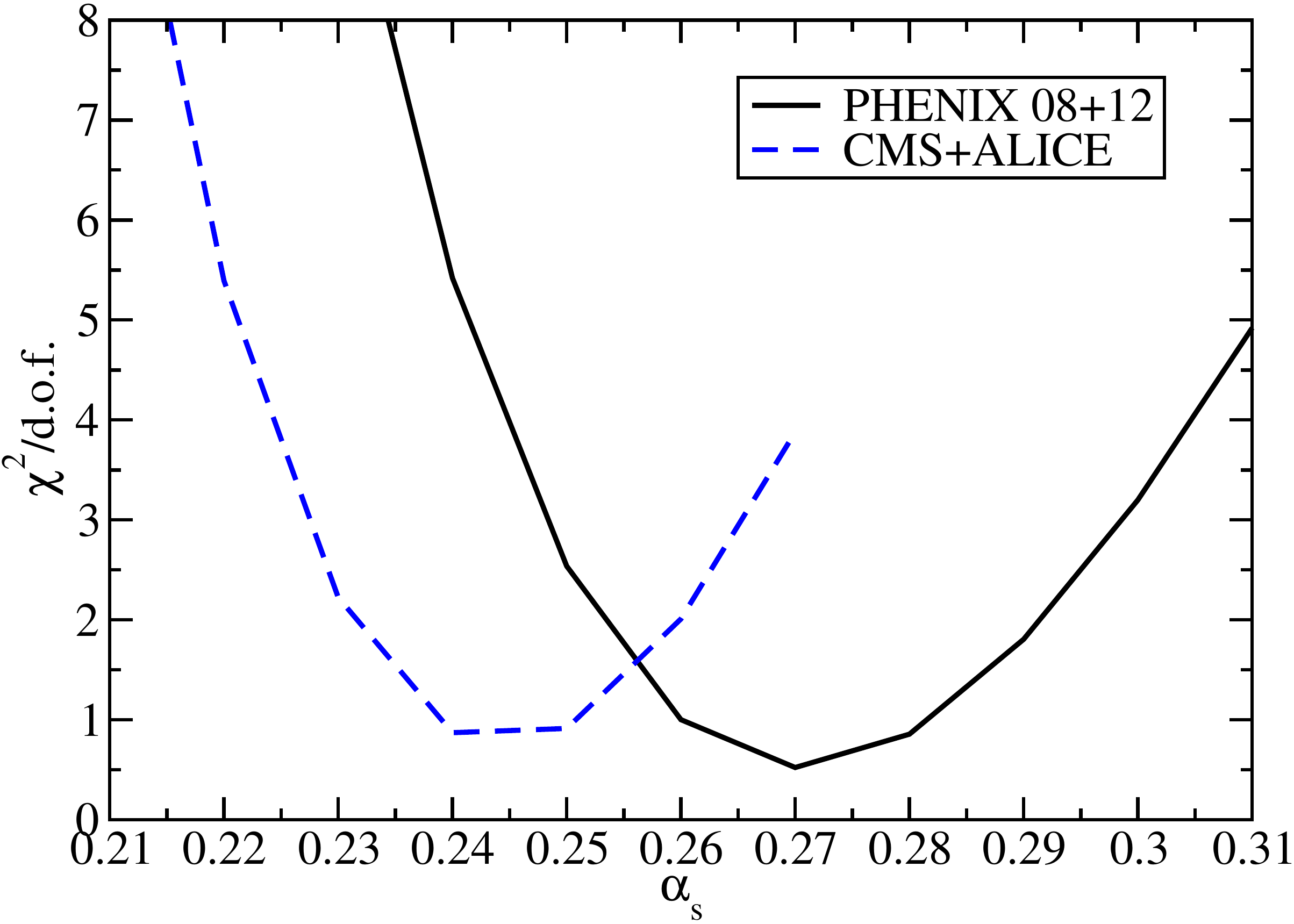}
\caption{(Color online) 
The $\chi^2$/d.o.f as a function of $\alpha_s$ from fitting to the PHENIX data \cite{Adare:2008qa,Adare:2012wg}  (combined 2008 and 2012 data set) at RHIC (solid) and combined ALICE \cite{Abelev:2012hxa} and CMS \cite{CMS:2012aa}  data at LHC (dashed) by the McGill-AMY model calculation of the nuclear suppression factor $R_{AA}(p_T)$ as shown in Fig.~\ref{mcgill-raa}.} 
\label{chisq_alphas}
\end{figure}

The above best $\alpha_s$ values are obtained from a $\chi^2$ fit, as shown in Fig. \ref{chisq_alphas}. Here the values of $\chi^2/{\rm d.o.f.}$ are plotted as a function of $\alpha_s$ for both RHIC and the LHC. For RHIC we use the data points above 5 GeV/$c$ for both 2008 and 2012 PHENIX data, for the LHC we use both CMS and ALICE data points with a momentum cut of $6$ GeV/$c$. 

\section{Jet transport parameter}

In order to compare medium properties extracted from phenomenological studies of jet quenching within different approaches to parton energy loss, we will focus on the value of quark jet transport parameter $\hat q$ either directly extracted or evaluated within each model with the model parameters constrained by the experimental data. As a first step, we will only consider data on the suppression factor of single inclusive hadron spectra $R_{AA}(p_T)$ at both RHIC and LHC. Within each model, $\hat q$ should be a function of both local temperature and jet energy which in turn varies along each jet propagation path. As a gauge of medium properties at its maximum density achieved in heavy-ion collisions, we will consider the value of $\hat q$ for a quark jet at the center of the most central  
A+A collisions at an initial time $\tau_0$ when hydrodynamic models are applied for the bulk evolution. For all the hydrodynamic models used in this paper with different approaches of parton energy loss, the initial time is set at $\tau_0=0.6$ fm/$c$ with initial temperature $T_0=346-373$ and 447-486 MeV at the center of the most central Au+Au collisions at $\sqrt{s}=200$ GeV/n at RHIC and Pb+Pb collisions at $\sqrt{s}=2.76$ TeV/n at LHC, respectively. 

\begin{figure}
\centering
\includegraphics[width=3.5in]{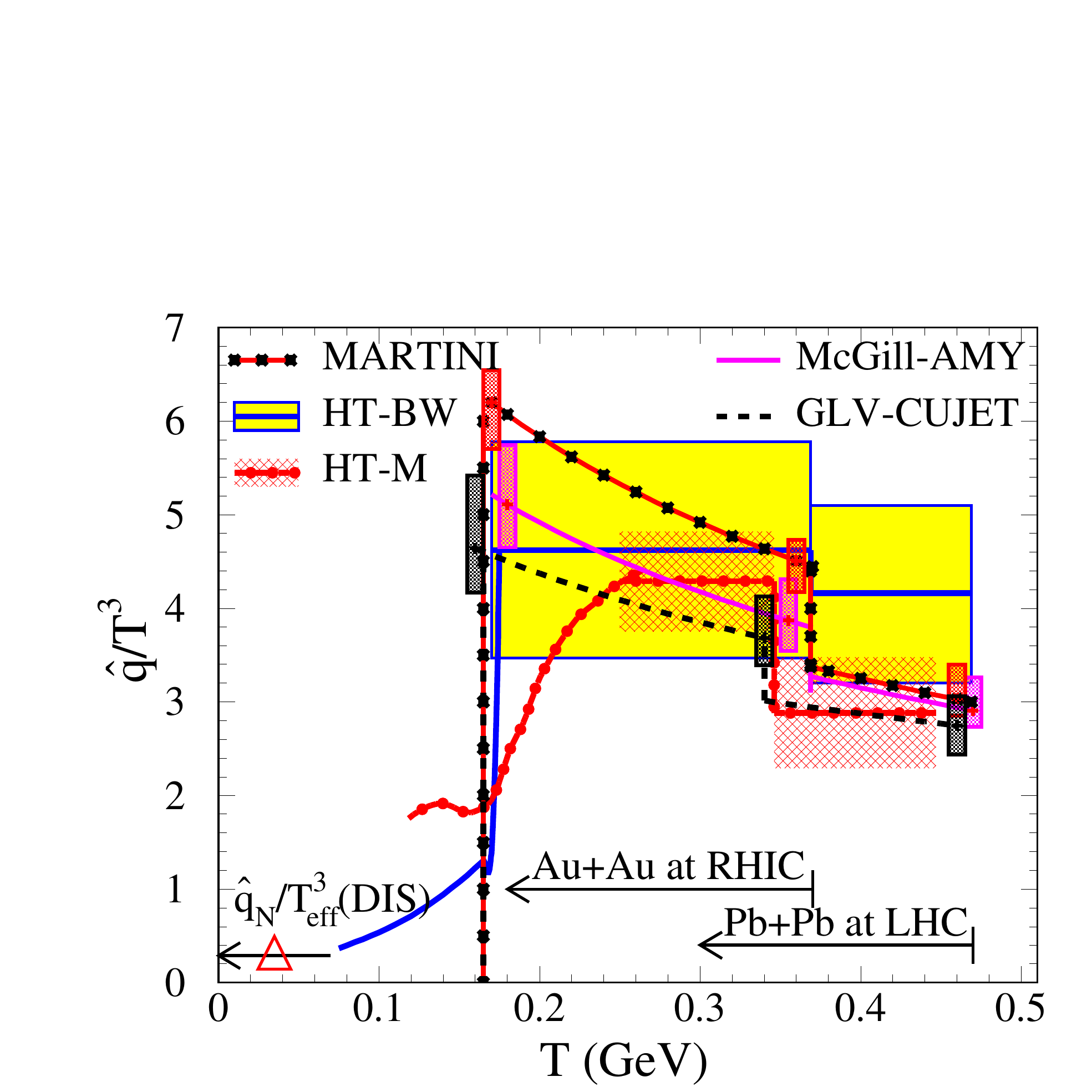}
\caption{\label{fig:raa_ht} (Color online) The assumed temperature dependence of the scaled jet transport parameter $\hat q/T^3$ in different jet quenching models for an initial quark jet with energy $E=10$ GeV.  Values of $\hat q$ at the center of the most central A+A collisions 
at an initial time $\tau_0=0.6$ fm/$c$ in HT-BW and HT-M models are extracted from fitting to experimental data on hadron suppression factor $R_{AA}$ at both RHIC and LHC. In GLV-CUJET, MARTINI and McGill-AMY model, it is calculated within the corresponding model with parameters constrained by experimental data  at RHIC and LHC. Errors from the fits are indicated by filled boxes at three separate temperatures at RHIC and LHC, respectively. The arrows indicate the range of temperatures at the center of the most central A+A collisions. The triangle indicates the 
value of $\hat q_N/T^3_{\rm eff}$ in cold nuclei from DIS experiments. } 
\label{fig:qhat}
\end{figure}

Shown in Fig.~\ref{fig:qhat} are the extracted or calculated values for $\hat q$ as a function of the initial temperature for a quark jet with initial energy $E=10$ GeV. For the GLV-CUJET model, $\hat q$ is calculated from one set of parameters with HTL screening mass and the maximum value of running coupling $\alpha_{\rm max}= 0.28$ for temperature up to $T=378$ MeV, and for another set with $\alpha_{\rm max}= 0.24$ for $378 \le T \le 486$ MeV. The difference in $\alpha_{\rm max}$ and the corresponding $\hat q$ in these two temperature regions can be considered part of the
theoretical uncertainties. 

Similarly, the values of $\hat q$ from the MARTINI and McGill-AMY models are calculated according to the leading order pQCD HTL formula in Eq.~(\ref{eq:localqhat}) with the two values of $\alpha_{\rm s}$ extracted from comparisons to the experimental data on $R_{AA}$ at RHIC and LHC, respectively.  The GLV, MARTINI and McGill-AMY models all assume zero parton energy loss and therefore zero  $\hat q$ in the hadronic phase.  In the HT-BW model, the fit to the experimental data gives $\hat q=1.3\pm 0.3 $ GeV$^2$/fm at temperatures reached in the most central Au+Au collisions at RHIC, and $2.2\pm 0.5$ GeV$^2$/fm at temperatures reached in the most central Pb+Pb collisions at LHC.  Values of $\hat q$ in the hadronic phase are assumed to be proportional to the hadron density in a hadron resonance gas model with the normalization in a cold nuclear matter determined by DIS data \cite{Deng:2009qb}. Values of $\hat q$ in the QGP phase are considered proportional to $T^3$ and the coefficient is determined by fitting to the experimental data on $R_{AA}$ at RHIC and LHC separately. In the HT-M model the procedure is similar except that $\hat q$ is assumed to be proportional to the local entropy density and its initial value is $\hat q=0.89\pm 0.11$ GeV$^2$/fm in the center of the most central Au+Au collisions at RHIC, and $\hat q=1.29\pm 0.27$ GeV$^2$/fm in the most central Pb+Pb collisions at LHC (note that the values of $\hat q$ extracted in Sec IV are for gluon jets
and therefore 9/4 times the corresponding values for quark jets).  For temperatures close to and below the QCD phase transition, $\hat q$ is assumed to follow the entropy density, and $\hat q/T^3$ shown in Fig.~\ref{fig:qhat} is calculated according to the parameterized EOS \cite{Huovinen:2009yb} that is used in the hydrodynamic evolution of the bulk medium.  In both HT approaches, no jet energy dependence of $\hat q$ is considered.

Considering the variation of the $\hat q$ values between the five different models studied here as theoretical uncertainties, one can extract its range of values as constrained by the measured suppression factors of single hadron spectra at RHIC and LHC as follows:
\begin{equation*}
\frac{\hat q}{T^3}\approx \left\{ 
\begin{array}{l}
4.6\pm 1.2 \qquad \text{at RHIC},\\
3.7 \pm 1.4 \qquad \text{at LHC},
\end{array}
\right.
\end{equation*}
at the highest temperatures reached in the most central Au+Au collisions at RHIC and Pb+Pb collisions at LHC. 
The corresponding absolute values for $\hat q$ for a 10 GeV quark jet are,
\begin{equation*}
\hat q \approx \left\{ 
\begin{array}{l}
1.2 \pm 0.3  \\
1.9 \pm 0.7 
\end{array}
 \;\; {\rm GeV}^2/{\rm fm} \;\; \text{at} \;\;
\begin{array}{l}
 \text{T=370 MeV},\\
 \text{T=470 MeV},
\end{array}
\right.
\end{equation*}
at an initial time $\tau_0=0.6$ fm/$c$. These values are very close to an early estimate \cite{Baier:1996sk} and are consistent with LO pQCD estimates, albeit with a somewhat surprisingly small value of the strong coupling constant as obtained in CUJET, MARTINI and McGill-AMY model.  The HT models assume that $\hat q$ is independent of jet energy in this study. CUJET, MARTINI and McGill-AMY model, on the other hand, should have a logarithmic energy dependence on the calculated $\hat q$ from the kinematic limit on the transverse momentum transfer in each elastic scattering, which also gives the logarithmic temperature dependence as seen in Fig.~\ref{fig:qhat}.

As a comparison, we also show in Fig.~\ref{fig:qhat} the value of $\hat q_N /T^3_{\rm eft}$ in cold nuclei as extracted from jet 
quenching in DIS \cite{Deng:2009qb} .  The value of $\hat q_N=0.02$ GeV$^2$/fm and an effective temperature of an ideal quark gas with 3 quarks within each nucleon at the nucleon density in a large nucleus are used. It is an order of magnitude smaller than that in A+A collisions at RHIC and LHC.

There are recent attempts \cite{Majumder:2012sh,Panero:2013pla} to calculate the jet transport parameter in lattice gauge theories.  A recent lattice calculation \cite{Panero:2013pla} found that the non-perturbative contribution from soft modes in the collision kernel can double the value of the NLO pQCD result for the jet transport parameter \cite{CaronHuot:2008ni}.  In the HT models such non-perturbative contributions could be included directly in the overall value of $\hat q$. They can also be included in the CUJET, MARTINI and McGill-AMY models by replacing the HTL thermal theory or screened potential model for parton scattering with parameterized collision kernels that include both perturbative and non-perturbative contributions.

One can also compare the above extracted values of $\hat q$ to other nonperturbative estimates. Using the AdS/CFT correspondence, the jet quenching parameter in a ${\cal N}=4$ supersymmetric Yang-Mills (SYM) plasma at the strong coupling limit can be calculated in leading order (LO) as \cite{Liu:2006ug}
\begin{equation}
\hat q_{\rm SYM}^{\rm LO}=\frac{\pi^{3/2}\Gamma(3/4)}{\Gamma(5/4)} \sqrt{\lambda} T^3_{\rm SYM},
\end{equation} 
where $\lambda=g_{\rm SYM}^2N_c$ is the 't Hooft coupling. To compare the SYM results to the extracted values of $\hat q$, one should take into account the different number of degrees of freedom in $N_c=3$ SYM and 3 flavor QCD. Since $\hat q$ is approximately proportional to the local entropy density (local gluon number density),  one can match the corresponding entropy density to obtain $3T_{\rm SYM}^3\approx T^3$. With a range of fixed values of $\alpha_s=0.22 - 0.31$ from CUJET, MARTINI and McGill-AMY fits, $\hat{q}^{\rm LO}_{\rm SYM}\approx 7.2 - 8.6$ is significantly above the range of $\hat q$ values in Fig.~\ref{fig:qhat} from 
 model fits to the experimental data on nuclear modification factors at RHIC and LHC.

Next to leading order (NLO) corrections to the above LO result \cite{Zhang:2012jd} due to world sheet fluctuations suggests,
\begin{equation}
\hat q_{\rm SYM}^{\rm  NLO}=\hat q_{\rm SYM}^{\rm LO}\left( 1- \frac{1.97}{\sqrt{\lambda}} \right).
\end{equation} 
One then gets
\begin{equation}
\frac{\hat q_{\rm SYM}^{\rm NLO}}{T^3} \approx 2.27 - 3.64 \qquad \text{for}\;\; \alpha_{\rm SYM}=0.22 - 0.31, \nonumber
\end{equation}
which falls within the range of $\hat q$ extracted from experimental data on $R_{AA}$ in Fig.~\ref{fig:qhat}. 

Other corrections of ${\cal O}(1/N_c)$ and higher orders in $1/\sqrt{\lambda}$ are also expected \cite{Noronha:2010zc,Ficnar:2013qxa,Ficnar:2013cka}. For example, part of the next-next-to-next leading order (NNNLO) corrections \cite{Armesto:2006zv} $-1.7552/\lambda^{3/2}$ is only about 5\% of the LO result. Other corrections at  next-to-next leading order (NNLO) correction $\propto \sim 1/\lambda$ are as yet undetermined.

\section{Conclusions and Outlook}

\begin{figure}
\centering
\includegraphics[width=3.5in]{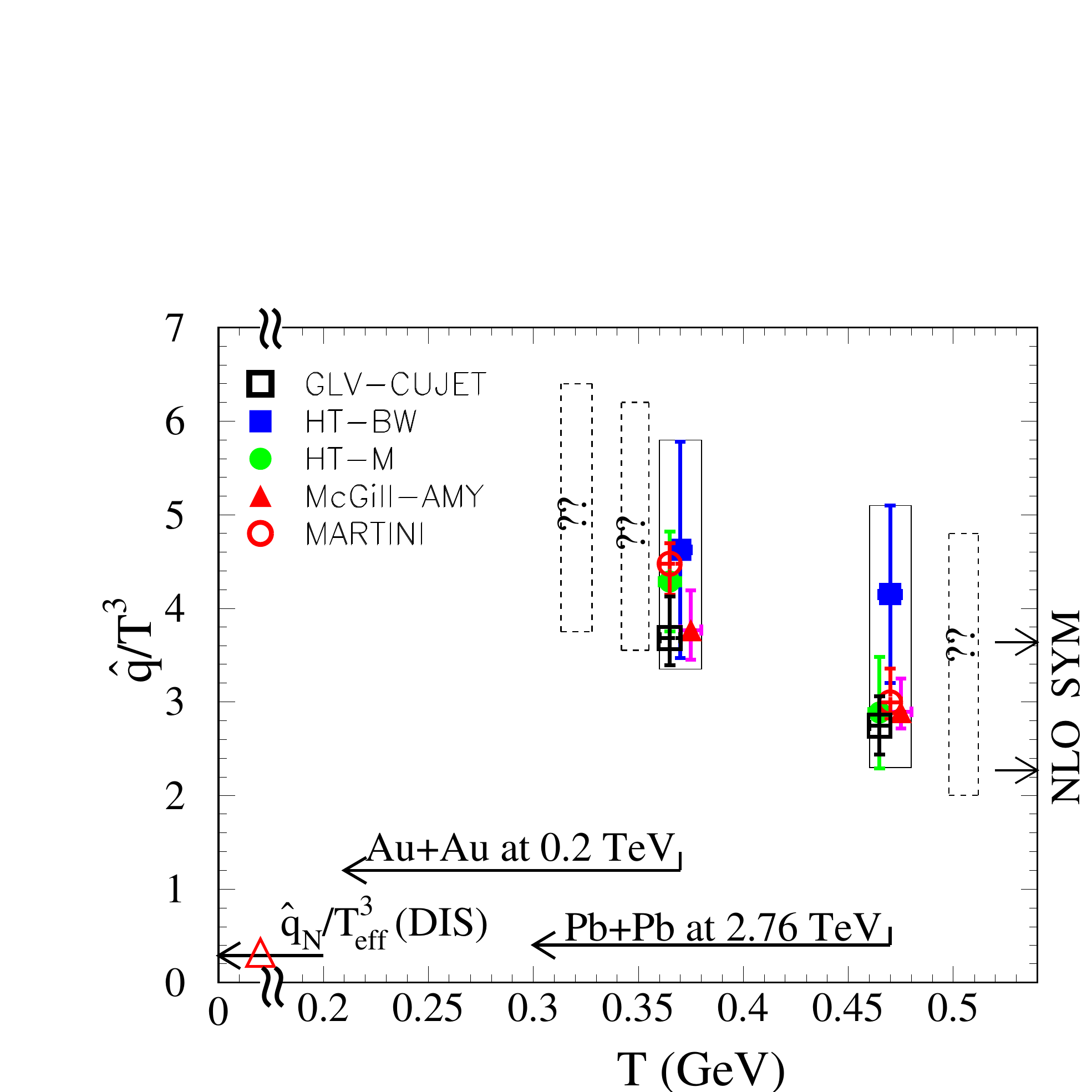}
\caption{\label{fig:raa_ht} (Color online) Values of scaled jet transport parameter $\hat q/T^3$ for an initial quark jet with energy $E=10$ GeV at the center of the most central A+A collisions at an initial time $\tau_0=0.6$ fm/$c$ constrained by experimental data on hadron suppression factor $R_{AA}$ at both RHIC and LHC. The dashed boxes indicate expected values in A+A collisions at $\sqrt{s}=0.063, 0.130$ and 5.5 TeV/n, assuming the initial entropy is proportional to the final measured charged hadron rapidity density \cite{Aamodt:2010pb}. The triangle indicates the value of $\hat q_N/T_{\rm eff}^3$ in cold nuclei from DIS experiments. Values of $\hat q_{\rm SYM}^{\rm NLO}/T^3$ from NLO SYM theory are indicated by two arrows on the right axis. } 
\label{fig:qhat2}
\end{figure}

We have carried out a survey study on the jet transport parameter extracted or calculated from five different approaches to the parton energy loss in a dense medium whose parameters are constrained by the experimental data on suppression factors of large transverse momentum hadron spectra in high-energy heavy-ion collisions at both RHIC and LHC. We find that new data from the LHC, combined with data from RHIC and advances in our understudying and modeling of jet quenching and bulk evolution,  provide much improved constraints on parton energy models. Compared to earlier efforts \cite{Bass:2008rv,Armesto:2009zi}, our present study significantly narrows down the variation of $\hat q$ extracted from
different jet quenching models and model implementations. The extracted value is surprisingly consistent with both pQCD and NLO AdS/CFT SYM results. The large range of $p_T$ covered by experimental data and the higher temperatures reached in the center of heavy-ion collisions at the LHC also allowed a first investigation of the jet energy and temperature dependence of the jet transport coefficient.

This is only a first step toward a systematic study of medium properties with hard probes constrained by the experimental data on a wide variety of observables that should include dihadron and gamma-hadron correlations, single jets, dijets and gamma-jets suppressions, azimuthal asymmetries, modification of jet profile and jet fragmentation functions. All of these studies should be carried out within a realistic model for jet quenching, hadronization and  bulk evolution that is also constrained by experimental data on bulk hadron spectra. This will require a full Monte Carlo simulation of the evolving jet shower in the expanding medium. With future precision and complementary complementary high statistics data from RHIC and LHC and theoretical advances in jet quenching and modeling of bulk evolution, it should be possible to further reduce the uncertainties in the determination of the jet transport parameters and to achieve a truly quantitative understanding of the QGP properties in high-energy heavy-ion collisions.

In Fig.~\ref{fig:qhat2}, we summarize our current results and
indicate needed future work by the JET collaboration
toward a quantitative mapping of the jet transport parameter $\hat q(E,T)$ over wider range of jet energy
and highest temperatures reached in the center of the most central A+A collisions. Experimental data on nuclear modification
factors  from lower energy ($\sqrt{s}=0.02 - 0.2$ TeV/n) at RHIC and future
higher energy (5.5 TeV/n) at LHC will likely help to further constrain
 models of jet-medium interactions. This will require extended effort with  both
viscous hydrodynamic calculations constrained by bulk observables as well jet quenching calculations
for light and heavy quark jets. An important challenging and open problem is to reconcile high $p_T$ jet 
azimuthal multipole moments, $v_n(p_T>10\;{\rm GeV}/c)$ with bulk ``flow'' moments in the
$p_T<2$ GeV/$c$ range. Concurrently, future improvements to theoretical calculations of the jet transport
parameter with both perturbative (NLO pQCD) and non-perturbative (lattice QCD and AdS/CET) methods 
will be required to reduce modeling uncertainties in jet quenching studies within the wide energy and temperature range
accesible at RHIC and LHC.

 \section*{Acknowledgement}
We would like to thank discussions with other members of the JET Collaboration.
This work was supported by the Director, Office of Energy
Research, Office of High Energy and Nuclear Physics, Division of Nuclear Physics, of the U.S. Department of 
Energy under Contract Nos. DE-AC02-05CH11231, DE-FG02-93ER40764, 
DE-AC02-98CH10886, DE-SC0004286
and  within the framework of the JET Collaboration, by the NSF under grant number PHY-1207918, 
by the Natural Sciences and Engineering Research Council of Canada,
by the Major State Basic Research Development Program in China (No. 2014CB845404) 
and by the National Natural Science Foundation of China under the grant Nos. 11221504, 11375072, 11175071 and 11035003.

\end{document}